%% file: Main.tex
\documentclass{ar-1col}
\usepackage[numbers,sort&compress]{natbib}
\usepackage{url}
\usepackage{aas_macros}
\jname{Annu. Rev. Nucl. Part. Sci.}
\jvol{71}
\jyear{2021}
\doi{10.1146/annurev-nucl-011921-061243}

\definecolor{purple}{rgb}{0.6, 0.4, 0.8}

\newcommand{\lsun}{{\rm L}_\odot}
\newcommand{\rsun}{{\rm R}_\odot}
\newcommand{\msun}{{\rm M}_\odot}
\newcommand{\tsun}{\tau_\odot}
\newcommand{\zxsun}{\left({\rm Z/X}\right)_\odot}

\begin{document}

\markboth{G. D. Orebi Gann, K. Zuber, D. Bemmerer, and A. Serenelli}{The Future of Solar Neutrinos}

\title{The Future of Solar Neutrinos}

\author{Gabriel D. Orebi Gann,$^{1,2}$  Kai Zuber$^3$, Daniel Bemmerer,$^4$ and Aldo Serenelli$^{5,6,7}$
\affil{$^1$Department of Physics, University of California, Berkeley, Berkeley, CA, USA, 94549}
\affil{$^2$Nuclear Science Division, Lawrence Berkeley National Laboratory, Berkeley, CA, USA, 94549}
\affil{$^3$Institute for Nuclear and Particle Physics, TU Dresden, Dresden, Germany, 01069}
\affil{$^4$Helmholtz-Zentrum Dresden-Rossendorf, 01328 Dresden, Germany}
\affil{$^5$Institute of Space Sciences (ICE, CSIC), Carrer de Can Magrans S/N, Bellaterra, Spain, E-08193}
\affil{$^6$Institut d'Estudis Espacials de Catalunya (IEEC), C/Gran Capita 2-4, Barcelona, Spain, E-08034}
\affil{$^7$Max Planck Institute for Astronomy, K\"onigstuhl 17, Heidelberg, Germany, D-69117}}
%
\begin{abstract}
In this article we review the current state of the field of solar neutrinos, including flavour oscillations, non-standard effects, solar models, cross section measurements, and the broad experimental program thus motivated and enabled.  We discuss the historical discoveries that contributed to current knowledge, and define critical open questions to be addressed in the next decade.
We discuss the state of the art of standard solar models, including uncertainties and problems related to the solar composition, and review experimental and model solar neutrino fluxes, including future prospects.
We review the state of the art of the nuclear reaction data relevant for solar fusion in the proton-proton chain and carbon-nitrogen-oxygen cycle.  Finally, we review the current and future experimental program that can address outstanding questions in this field.
\end{abstract}
\begin{keywords}
keywords, separated by comma, no full stop, lowercase, solar neutrino, solar fusion, proton-proton-chain, carbon-nitrogen-oxygen cycle, astrophysical S-factor, thermonuclear reaction rate
\end{keywords}
\maketitle
\tableofcontents

\input{intro}
\input{models1}

\input{nuclear}

\input{models2}

\input{solarphysics}

\input{experiments}


\section{CONCLUDING REMARKS}\label{s:conc}

While recent decades have offered tremendous advances in solar neutrinos, across the fields of astro-, nuclear, and particle physics, many lingering mysteries remain.  

The solar abundance problem remains open.  Whether  spectroscopic abundances, radiative opacity, or non-standard modifications to solar models lie at its core remains to be seen. While progress is being made on all these fronts, independent measurements of solar abundances, ideally free from model dependencies, are needed. Solar neutrinos from the CNO cycle are a unique opportunity in this regard and, in addition, offer a direct view on the pristine composition of our solar system. Borexino \cite{2020Natur.587..577B} has provided the first detection of these neutrinos, opening the road for future neutrino experiments.  Given  its size and depth, SNO+ could improve on the precision of the CNO neutrino flux measurement if sufficient background reduction can be achieved.  Perhaps the best prospects for enhanced precision of future measurements beyond SNO+ lie in the concept of a hybrid detector, offering directional sensitivity via Cherenkov detection in a low-threshold scintillation detector.  

A precise measurement of pp neutrinos, another milestone for experimental solar neutrino physics, is needed to establish tight limits on the origin of the solar energy, and set limits on non-standard energy channels.  The best prospects for this may lie with multi-purpose detectors: the low thresholds of liquid noble gas detectors, built for rare event searches, may yield excellent precision on pp neutrinos with sufficient control of background sources.  Alternatively, large organic scintillator detectors may have good sensitivity, given sufficiently low levels of $^{14}$C.

Low-energy neutrinos are best addressed in large, low-threshold scintillator detectors.  Precision measurements of the $^7$Be and $^8$B fluxes offer insights into the solar core temperature, and non-standard effects.  A spectral measurement of $^8$B neutrinos in the transition region between matter- and vacuum-dominated oscillation offers a unique chance to probe the details of the interaction of neutrinos with matter, with sensitivity to a range of potential non-standard effects.  At the other end of the solar neutrino spectrum, large water Cherenkov detectors such as Hyper-K, and the DUNE LAr experiment offer the potential for a significant observation of the day-night asymmetry, thus confirming the regeneration of electron-flavour neutrinos in the earth at night, and further constraining oscillation parameter values.  A first measurement of hep neutrinos would complete the picture of solar neutrino fluxes.  A precision measurement of $\Delta$m$^2_{12}$ could resolve the current small tension between reactor and solar data.  New data will come from Super-K+Gd, JUNO, SNO+, and other projects.

The nuclear reaction data for the Sun
has yet to match the precision given by the $^7$Be and $^8$B neutrino fluxes. In order to address this imbalance,  further experimental and theoretical efforts on nuclear reactions and a consolidated new evaluation of existing results are called for. The same is true for the CNO neutrino flux: the precision of the controlling CNO nuclear reaction data must be improved, using experiment, theory, and evaluations. 
As more data are gathered and new experiments come online, the next decade may offer insights into several exciting open questions in this field, with the potential to inform our understanding of solar evolution, as well as neutrino properties.

\begin{summary}[SUMMARY POINTS]
\begin{enumerate}
\item The solar abundance problem still awaits a solution. 
\item Current solar neutrino fluxes favor a solar core with temperature profile comparable to that in HZ SSMs, in agreement with helioseismic inferences of the solar sound speed and other helioseismic diagnostics.
\item A measurement of the flux of neutrinos from the CNO cycle allows a derivation of the total abundance of C and N in the solar core, almost independently of solar model uncertainties.
\item Nuclear energy as the origin of solar luminosity is understood to 7\% (1$\sigma$).
\item The nuclear reactions important for solar neutrinos have all been qualitatively identified, and their quantitative knowledge is precise to 5-20\%, depending on the reaction considered. 
\item Neutrinos from all solar neutrino production branches have been observed, except the hep-neutrinos.
\item The current knowledge of the electron survival probability curve allows room for physics beyond the standard model.
\item The future experimental program will likely rely on multi-purpose detectors, with broad programs.
\item The strongest solar neutrino program will leverage results from a wide range of experiments, with complementary target materials, detection technology, and interaction types.
\end{enumerate}
\end{summary}

\begin{issues}[FUTURE ISSUES]
\begin{enumerate}
\item Precise measurement of neutrinos from the CNO cycle would yield an independent indication of solar core composition. A 10\% measurement would give results comparable in precision to spectroscopic techniques, but almost free of model systematics.
\item Measurements of the electron neutrino survival probability in the energy region 2--5 MeV would be desirable to constrain and probe new physics.
\item Improved precision and additional real-time measurements of pp solar neutrinos would offer new opportunities to study the Sun.  
\item Observation of  hep neutrinos would add the final branch of the fusion chains to current measurements.
\item New laboratory and theoretical work is needed to improve the precision of solar fusion cross sections: for the pp-chain $^3$He($\alpha,\gamma$)$^7$Be and $^7$Be($p,\gamma$)$^8$B to 3\% and for the CNO cycle $^{14}$N($p,\gamma$)$^{15}$O to 5\%. 
\end{enumerate}
\end{issues}

\section*{DISCLOSURE STATEMENT}
The authors are not aware of any affiliations, memberships, funding, or financial holdings that
might be perceived as affecting the objectivity of this review. 

\section*{ACKNOWLEDGMENTS}
The authors wish to thank Wick Haxton for his expertise and numerous insights, and Steve Biller, Mark Chen, Josh Klein, Sean Paling, and Nigel Smith for fruitful discussions and input.
Support is gratefully acknowledged from: (Germany) DFG (BE 4100/4-1, ZU 123/21-1) and the COST Association (ChETEC, CA16117);  (Spain) the Spanish Government through the MICINN grant PRPPID2019-108709GB-I00; (USA) the Director, Office of Science, of the U.S. Department of Energy under Contract No. DE-AC02-05CH11231, and the U.S. Department of Energy, Office of Science, Office of Nuclear Physics, under Award Number  DE-SC0018987.

\bibliography{References}
\bibliographystyle{ar-style5}
\end{document}

%% file: intro.tex
\section{INTRODUCTION AND HISTORICAL OVERVIEW}\label{s:intro}
\begin{figure}
\includegraphics[width=0.8\textwidth]{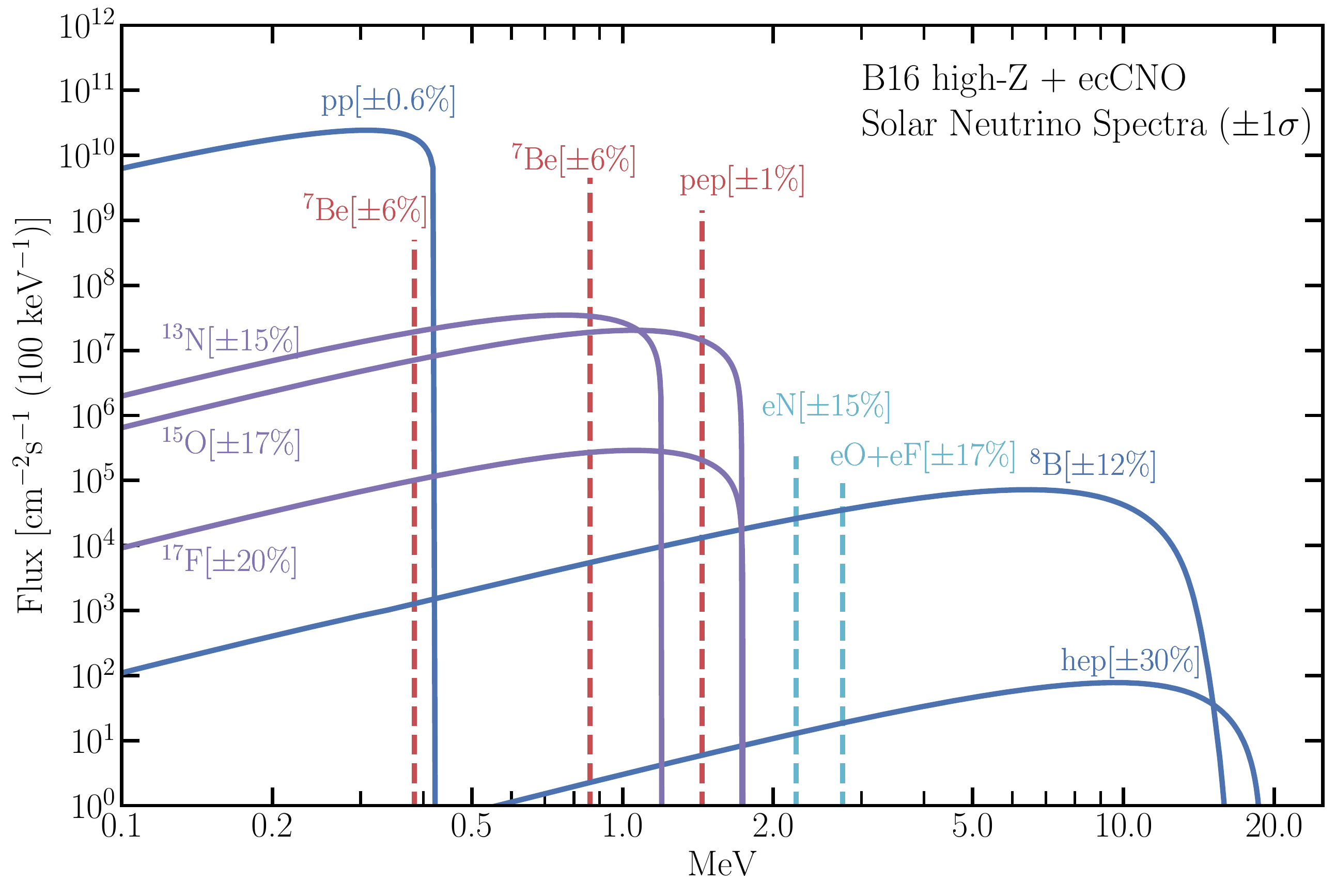}
\caption{Spectrum of neutrino fluxes from different nuclear reactions (see Sec.~\ref{sec:Nuclear}), using results from \cite{vinyoles:2017}. Neutrino fluxes from electron capture reactions are given in $\hbox{cm}^{-2}\hbox{s}^{-1}$. \label{fig:nuspectrum}}
\end{figure}

For over 100 years, our understanding of the Sun, and its energy production, was  based on thermodynamic and geological arguments. The modern era was ushered in by an experimental revolution,  starting with the discovery of radioactivity and the Rutherford experiment \cite{Rutherford1902}.
In 1920, Lord Eddington suggested that a fusion process might be sufficient for energy generation in stars. The discovery of the tunnel effect by Gamow led Atkinson and Houtermans to the conclusion that high energy protons (based on the interplay of Maxwell-Boltzmann distribution and tunneling effect) and small nuclear charge would be favourable to allow fusion processes. This led to two fundamental reaction schemes -- the pp chain and the CNO cycle -- with a mutual equation \cite{Weizsaecker38-PZ,1939PhRv...55..103B}:
\begin{equation}
    4 p \rightarrow \alpha + 2\nu_e + 26.73 \mbox{MeV}, 
\end{equation}
where neutrinos carry away a small fraction of the total energy available.  
Neutrinos are produced by several reactions, each giving rise to a characteristic energy distribution, or spectrum. The contributions of different reactions to the solar neutrino spectrum are illustrated in Figure~\ref{fig:nuspectrum}.  All neutrinos produced in these cycles are created in the electron flavour.
Observation of these neutrinos can offer insights into both the Sun, and neutrino properties.  A decades-long campaign has yielded regime-altering results, along with two Nobel Prizes.  Yet a number of mysteries remain.

In 1946, Pontecorvo suggested to use nuclear transitions for neutrino detection. He proposed the use of the reaction $^{37}$Cl$\rightarrow ^{37}$Ar for a radiochemical experiment. The threshold for neutrino capture  for this reaction is 814~keV. Collecting the produced $^{37}$Ar atoms using small proportional counters for the detection of the electron capture of $^{37}$Ar served as signal. Based on the unique physical and chemical properties, this proposal developed into the successful Homestake experiment with the first observation of solar neutrinos \cite{1968PhRvL..20.1205D,1998ApJ...496..505C}.  This achievement earned Ray Davis Jr. the 2002 Nobel Prize in Physics.
The experiment surprisingly found a deficit in the solar neutrino flux, detecting only 1/3 of the expected signal. 
The energy threshold of the chlorine reaction used in the Homestake experiment meant it was not sensitive to pp-neutrinos, which sit very low in the spectrum (Fig.~\ref{fig:nuspectrum}).
Access to these pp neutrinos was achieved using similar radiochemical techniques, with the reaction $^{71}$Ga $\rightarrow ^{71}$Ge, which benefits from a Q-value of only 233 keV. These measurements were performed by GALLEX \cite{1992PhLB..285..376A,1999127,2010PhLB..685...47K} and later GNO \cite{2005PhLB..616..174G} in the Italian Laboratori Nazionali del Gran Sasso, and SAGE \cite{PhysRevC.80.015807,2019sone.conf...29G} in the Russian Baksan laboratory, each of which showed a deficit in the observed solar neutrino flux. 

Around a similar time, solar neutrinos were observed by the water Cherenkov experiment Kamiokande~\cite{PhysRevLett.63.16,PhysRevLett.77.1683} via elastic scattering (ES) on electrons. This result confirmed the presence of a deficit with an independent method, and achieved the first real-time detection of solar neutrinos.  The directional nature of the ES signal was critical in confirming that the observed neutrinos were coming from the Sun. 
The experiment was upgraded to Super-Kamiokande, a massive detector with a broad physics program, that has operated successfully for several decades. The latest solar neutrino results can be found in~\cite{2016PhRvD..94e2010A}. 
The threshold for water Cherenkov detection is several MeV, making these experiments sensitive primarily to $^8$B neutrinos. The detected flux was approximately one half that expected based on solar models.  

This deficit is what came to be known as the ``solar neutrino problem'', and a clear energy dependence was observed across the different experimental results.
Many solutions were proposed, with foundations in astrophysics, nuclear physics, and particle physics. A favourite explanation was neutrino oscillation, the periodic change among the three neutrino flavors. Furthermore, it was recognised that the behaviour of neutrinos in matter is modified by the electron density and, in particular, the adiabatic change of electron density to which the neutrino is subjected as it propagates out from the core of the Sun. This causes an additional flavour-changing effect, due to the presence of charged current (CC) reactions, as well as neutral currents (NC) for the electron flavour neutrinos initially produced in fusion reactions~\cite{1978PhRvD..17.2369W,1985YaFiz..42.1441M}.  
The pattern of fluxes across these experiments motivated a new generation of projects, seeking to resolve the solar neutrino problem via improved precision,  direct detection, and enhanced flavour information.

The Sudbury Neutrino Observatory (SNO) was built to resolve the solar neutrino problem \cite{PhysRevLett.55.1534,2000NIMPA.449..172B}. Based on heavy water (D$_2$O), this experiment offered both a NC reaction and CC reaction for neutrino detection, as well as the ES used in light water Cherenkov detectors.
This allowed the SNO collaboration to detect both the pure electron flavour (via CC) and the total flux (via NC) of solar neutrinos, thus demonstrating unequivocally that the measured total solar flux agreed with solar model calculations \cite{snocc,2004PhRvL..92r1301A}, and that the deficit was due to neutrino flavour change. In 2015, Arthur McDonald and Takaaki Kajita were co-awarded the Nobel Prize in Physics ``for the discovery of neutrino oscillations, which shows that neutrinos have mass''. Results from SNO remain the only model-independent measurement of the total solar neutrino flux. 

The Borexino liquid scintillator experiment at LNGS \cite{2009NIMPA.600..568B} was constructed to offer improved spectral precision, and sensitivity to lower energy neutrinos than can be achieved with water detectors.  
Of particular note is the astonishingly low background achieved in this detector due to unprecedented levels of cleanliness and thermal control.  Such backgrounds typically arise from radioactive contaminants such as the naturally-occurring decay chains of $^{238}$U and $^{232}$Th, $^{14}$C and $^{40}$K. 
By mitigating numerous potential sources of such background, the Borexino collaboration has been able to complete spectroscopic measurements of all the pp chain reactions except the hep-neutrinos~\cite{borexino:pp2, borexino:pp3}, and  also recently provided the first observation of CNO neutrinos \cite{2020Natur.587..577B}. 
\begin{marginnote}[]
\entry{LNGS}\\
{Laboratorio Nazionale del Gran Sasso}
\end{marginnote}

The KamLAND liquid scintillator experiment in Japan used anti-neutrinos from nuclear reactors to demonstrate that the observed flavour change was in fact due to oscillation~\cite{2003PhRvL..90b1802E}.   
These results significantly constrained the parameter space for solar neutrino oscillation, and remain the only terrestrial measurement of these parameters.

A number of exciting questions remain in this field.  In the remainder of this article we describe the current status and prospects for advancement.   In Section~\ref{s:models1} we describe current solar models.  In Section~\ref{sec:Nuclear} we discuss our current understanding of nuclear reactions and cross section measurements needed for future improvements. In Section~\ref{s:models2} we present the current best knowledge of solar neutrino fluxes and the dominant uncertainties.  In Section~\ref{s:physics} we discuss
the details of solar neutrino physics, including potential non-standard effects that could affect oscillation behaviour, and in Section~\ref{s:expt} we describe the broad experimental program, including multi-purpose detectors with sensitivity to solar neutrinos, with an outlook intended to cover the next decade.  In Section~\ref{s:conc} we conclude with a summary and discussion of future prospects.

%% file: models1.tex
\section{SOLAR MODELS \label{s:models1}}

Two qualitatively different classes of solar models are used to study solar interior properties:  seismic and  evolutionary models. Seismic models use helioseismic data, primarily the frequencies of the solar pressure modes (p-modes), to reconstruct the solar internal structure using inversion methods \cite{shibahashi:1996,couvidat:2003,buldgen:2020}. They are structural models in the sense that they represent a static picture of the Sun and do not consider its evolution. This class of models is useful to understand the caveats present in solar models, e.g. missing physical processes, because they reproduce, by construction, the mechanical structure of the Sun. The accuracy and precision of these models is hampered in the innermost solar core (R$\leq 0.1\rsun$), the region where most of the $^7$Be, and almost the totality of $^8$B and CNO neutrinos are produced. The majority of p-modes do not reach those regions, as the internal boundary of their propagation cavity is located outside that region \cite{jcd:2002,aerts:2010}. 
Evolutionary models, on the other hand, follow the evolution of the stellar model since its formation to its present-day age $\tsun=4.57\,\hbox{Gyr}$ by integrating spatially and in time the equations of stellar structure and evolution. The minimum set of requirements imposed on this class of models is that at $\tsun$ the model has $1\,\msun$, and it reproduces the solar luminosity $\lsun$, radius $\rsun$ and solar surface (photospheric) composition. Evolutionary models are defined by the physical processes included in the equations of stellar evolution. Standard solar models (SSM) \cite{bahcall:1995,modelS,scilla:1997,turck:1999,bahcall:2002,serenelli:2009} include physics that is considered standard in stellar evolution, and that the Sun has lost a negligible amount of mass during its lifetime. Additional processes such as dynamical ones induced by rotation, magnetic field, gravity waves, and others can also be included and give rise to non-standard solar models (non-SSM) \cite{pinsonneault:1989,schlattl:1999,turck:2010,guzik:2010,zhang:2019}. A recent and very thorough review on SSMs and non-SSMs is \cite{jcd:2020}. 

\subsection{Solar composition and the solar abundance problem}

The chemical composition of the solar photosphere, in particular for elements heavier than helium, is determined with spectroscopic techniques, which require detailed solar atmosphere models, atomic data and treatment of spectral line formation under non local thermodynamic equilibrium \cite{asplund:2005,2009ARA&A..47..481A,bergemann:2014}. Determination of spectroscopic chemical abundances involves a strongly model dependent procedure, including subjective choices.  The abundance of refractory elements can be determined, relative to a reference element -typically silicon- from ancient meteorites known as CI carbonaceous chondrites \cite{lodders:2003}. The meteoritic scale can then be placed on the photospheric scale using again one or several refractory elements with reliable photospheric measurements as anchor points \cite{lodders:2003}. The \emph{solar mixture} is a critical constraint for solar models, and individual elements are relevant as far as they play a significant role in the radiative opacity in the solar interior or in the operation of the CNO cycle (e.g. the primordial solar nitrogen abundance is relevant for the CNO cycle, but has a very minor role in the radiative opacity in the solar interior). Solar mixtures are usually characterized by the photospheric total metal to hydrogen mass ratio, $\zxsun$, but the detailed abundance pattern is relevant for solar models.
\begin{marginnote}[]
\entry{Solar mixture}{Relative distribution of solar photospheric abundances for all elements heavier than helium (metals).}
\end{marginnote}

\begin{figure}
\includegraphics[width=\textwidth]{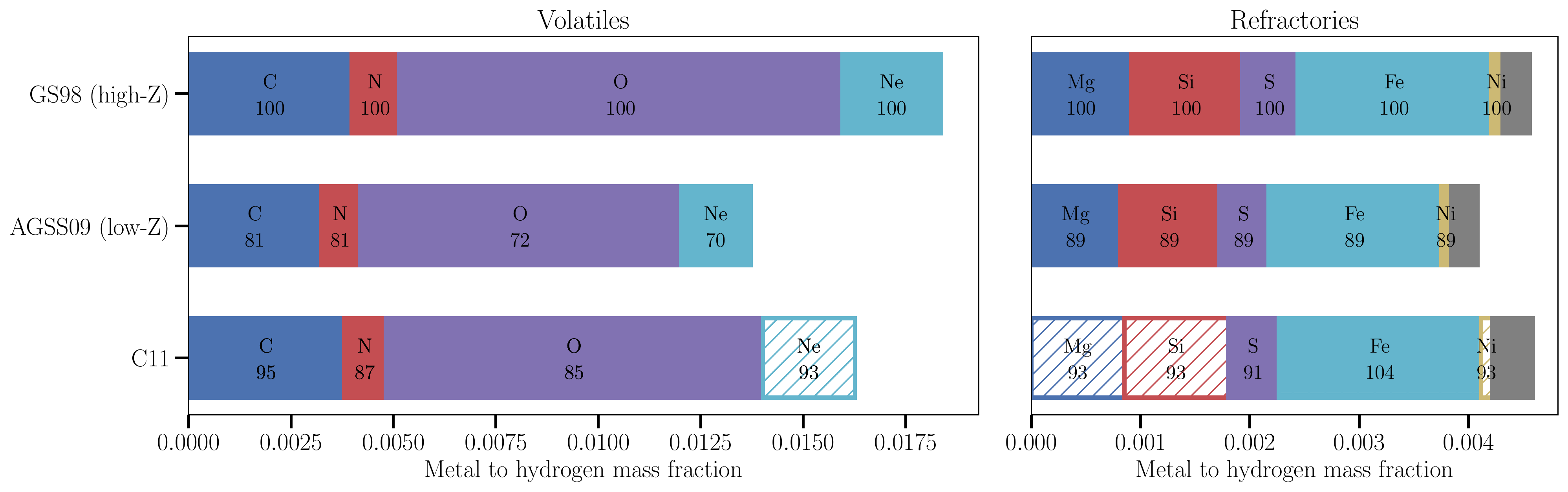}
\caption{Summary of three different solar mixtures. Left panel:photospheric metal to hydrogen mass ratio for volatile elements, which constitute most of the solar metallicity. Right panel: same but for most important refractory elements that contribute to the radiative opacity in the solar interior. Note the different scales of the plots. Numbers indicate the abundance in percentage of a given element with respect to its GS98 abundance (100\% by construction). Hatched areas denote elements not determined by C11, but taken from \cite{lodders:2009}.
 \label{fig:composition}}
\end{figure}

The development of three-dimensional radiation hydrodynamics models, improved atomic data and non-local thermodynamic equilibrium (NLTE) modeling of line formation led to a complete revision of the solar mixture \cite{2009ARA&A..47..481A}. In this review we refer to these abundances as AGSS09, or low-Z, in the astrophysical nomenclature. Partial revisions with  similar techniques are also available \cite{caffau:2011} (C11). Other hybrid solar mixtures, combining more heterogeneous data sets are also available \cite{lodders:2003,lodders:2009}.  The largest variations with respect to older solar mixtures \cite{1998SSRv...85..161G} (GS98), predating the aforementioned developments  in spectroscopy, occurred for volatile elements, particularly C, N, O, and Ne. We refer to the latter work as GS98, or high-Z. A comparison of the solar abundances of the most important elements for solar modeling is presented in Figure~\ref{fig:composition} for the GS98, AGSS09, and C11 solar mixtures. 
Typical uncertainties for each of the elements are in the range 10-12\% for volatile elements and $<5\%$ for refractory elements. The total photospheric metal to hydrogen ratio is $\zxsun= 0.0229, 0.0178, 0.0209$ respectively for the three mixtures.

\begin{figure}
    \centering
    \includegraphics[width=0.8\textwidth]{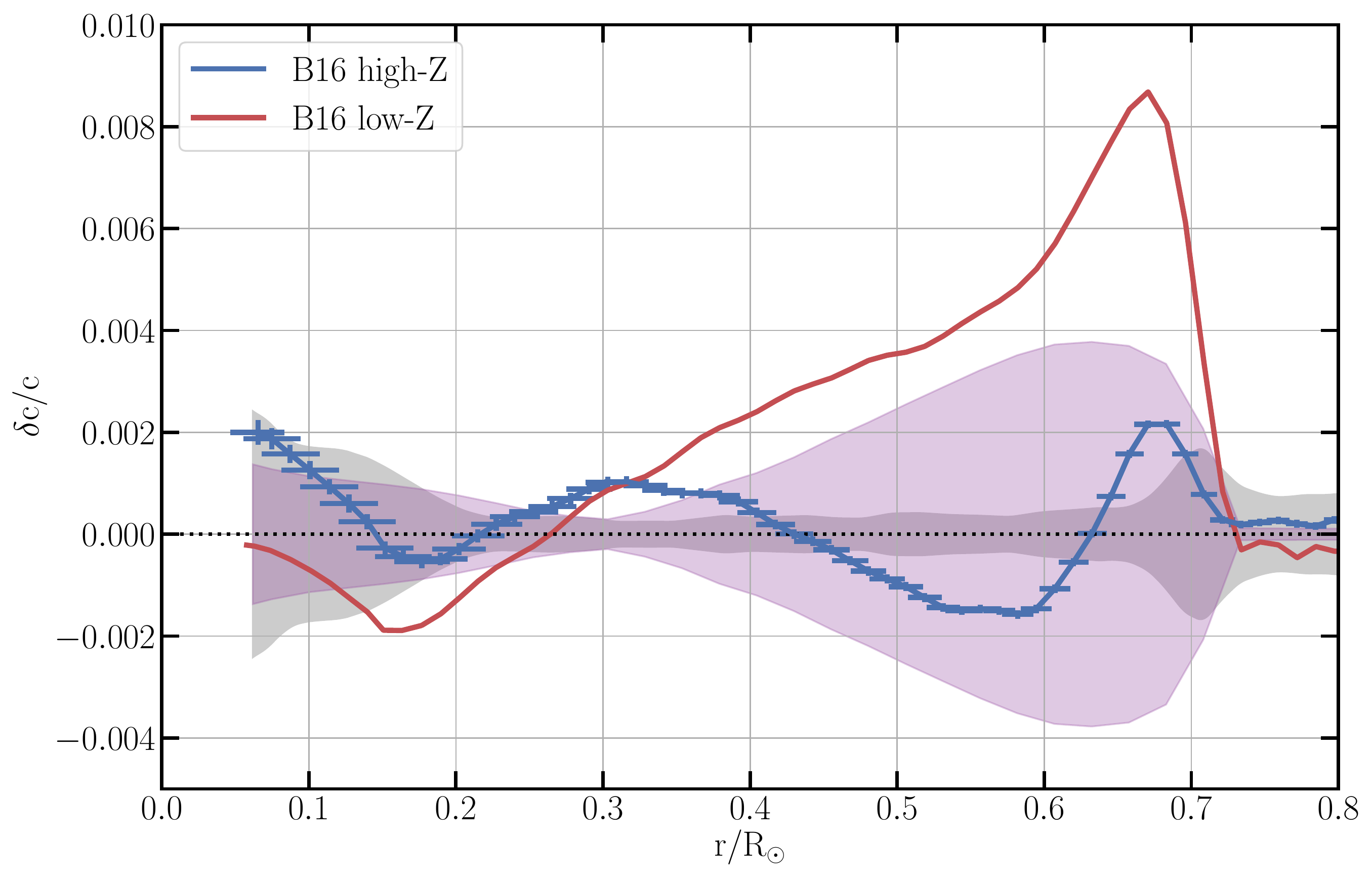}
    \caption{Fractional sound speed difference in the sense (Sun - model)/model for SSMs based on high-Z and low-Z solar abundance mixtures. Error bars denote uncertainties due to measurement uncertainties in the solar acoustic oscillation frequencies (vertical direction) and size of kernels in the inversion (horizontal direction). Shaded areas represent model (pink) and inversion technique (grey) 1$\sigma$ errors \cite{vinyoles:2017}. 
    \label{fig:csound}}
\end{figure}

The adoption of a low-Z solar mixture leads to SSMs  that do not reproduce the solar internal structure when compared to results from helioseismology. This is a result of the reduced radiative opacity in solar models with lower metallicity, primarily but not only due to the reduction in oxygen and neon abundances. The radiative opacity regulates energy transport in the solar interior, and therefore the mechanical (pressure-density) structure to which solar acoustic oscillations are sensitive. This is at odds with results stemming from SSMs based on the older, higher, solar metallicity. SSMs that use a high-Z solar mixture, although not perfect, reproduce helioseismic results much more satisfactorily \cite{modelS,bahcall:2005,jcd:2009,serenelli:2009,turck:2010,guzik:2010,vinyoles:2017}. This is best seen in Figure~\ref{fig:csound}, which shows the fractional sound speed difference between SSMs with high-Z and low-Z compositions, and the Sun. The discrepancy between low-Z solar models and helioseismic  inferences on the solar interior, the \emph{solar abundance problem}, has been the subject of numerous works, starting around 2004, \cite{bahcall:2004, montalban:2004, bahcall:2005, antia:2005, delahaye:2006,castro:2007,guzik:2010}, including some comprehensive reviews \citep{basu:2008,jcd:2020}. 

Quantifying the disagreement between SSMs and helioseismic and neutrino data is difficult. Several helioseismic diagnostics can be used: depth of the convective envelope, surface helium abundance, structural inversion of sound speed, density, adiabatic index $\Gamma_1$, among others \cite{jcd:2002,basu:2008,buldgen:2020}. However,  correlations among them, and among different regions in the Sun, are not properly quantified in the literature. Moreover, systematic effects related to different methodologies for carrying out helioseismic inversions are not usually addressed \cite{scilla:1997, vinyoles:2017}. The most recent and comprehensive efforts, which rely on sound speed and density inversions and experimental determination of solar neutrino fluxes, yield 0.9$\sigma$ and 3.0$\sigma$ agreement level between model and data for the GS98- and AGSS09-based SSMs, respectively \cite{villante:2014,vinyoles:2017,song:2018}.

Several tentative solutions have been put forward to solve the solar abundance problem, both within and outside the framework of SSMs. These include, among others: increase of gravitational settling to produce an internal metal-rich and surface metal-poor Sun; increased neon abundance; accretion of metal-poor material from the protoplanetary disk; and solar models with strong mass loss \cite{montalban:2004,bahcall:2005b,guzik:2006,castro:2007,haxton:2008,guzik:2010,turck:2011}. No satisfactory solution has been found so far. This is partly due to the fact that restoring agreement in a given helioseismic observable sometimes requires modifications in the solar model inputs opposite to those needed to restore agreement in another observable. The most illustrative example is given by the depth of the convective zone and surface helium abundance \cite{delahaye:2006, serenelli:2011}. More contrived models, combining several modifications to solar models, have been also proposed \cite{zhang:2019}, and can lead to better agreement, although after several ad-hoc assumptions. The topic of modified SSMs and non-SSMs has been reviewed extensively in  recent literature \cite{bergemann:2014,basu:2016,jcd:2020}. 

Modern spectroscopic analysis methods are qualitatively superior to older ones. But the solar abundance problem draws attention to the question of which is the correct photospheric solar composition and, by extension, that of the solar interior. Alternatively, does the current level of disagreement between low-Z SSMs and helioseismology set the best possible accuracy with which standard solar models can reproduce the solar internal structure? Independent determinations of the solar metal content are not only desirable, but fundamental. They hold the key to the our detailed understanding of the Sun and  all other stars. 

The main difficulty in such an independent measurement lies in the fact that  metals affect the solar structure only indirectly, e.g. through radiative opacities or the equation of state. Helioseismic diagnostics such as the sound speed profile, depth of convective zone and surface helium abundance of low-Z SSMs can be made to match helioseismic results by a well-chosen increase in radiative opacities, i.e. the degeneracy between metals and atomic opacities is almost complete \cite{jcd:2009}. Alternatively, helioseismic techniques have been used to determine the metallicity of the solar envelope, which should match that of the photosphere. Whereas some works find results consistent with high-Z solar mixtures \cite{lin:2007} others, more recently, have found metallicity values consistent with those of the AGSS09 mixture, and even lower \cite{vorontsov:2013,buldgen:2017}. Caution is in order, as the helioseismic signal induced by metals is feeble and entangled with that of helium; results rely on the accuracy of the equation of state; and systematic uncertainties are difficult to address. On the positive side, such measurements correspond to the adiabatic region of the convective envelope and are therefore independent of the radiative opacities, which are a critical but poorly known input in SSMs (Sec.~\ref{subsec:uncertainties}).

\begin{marginnote}[]
\entry{Metals}{
 affect solar interior models through the radiative opacity. }
\entry{Low-Z solar models} {show a global 3$\sigma$ disagreement with helioseismic data.} \entry{High-Z models}{differ by only 0.9$\sigma$.}
\entry{The solar abundance problem}{ arises due to a reduction by 30 -- 40\% of the inferred C, N, O, and Ne abundances from novel spectroscopic analysis methods, and represents a conflict between state-of-the-art spectroscopic methods and solar structure models. }
\end{marginnote}

\subsection{Uncertainties in solar models \label{subsec:uncertainties}}

SSMs establish a well defined framework in which only three free parameters are adjusted to reproduce the observational constraints: the initial hydrogen and helium abundances and one parameter associated with the treatment of convection. On the other hand, all other physical processes that are included in non-SSMs  require additional free parameters that are tuned either using observational data (e.g. lithium abundance \cite{schlattl:1999}, thus removing the model's capability to make testable predictions of this quantity); on hydrodynamic simulations carried out in physical regimes far from those of the solar interior \cite{zhang:2019}; or simply on the best judgment of the researcher (e.g. composition of accreted material \cite{serenelli:2011}). In this regard, non-SSMs are to some extent phenomenological evolutionary models that are built to explore possible missing physics in SSMs. But quantification of uncertainties is then restricted to SSMs. 

Recent work quantifying model uncertainties for neutrino fluxes and helioseismic diagnostics includes \cite{serenelli:2013, villante:2014, vinyoles:2017, villante:2021}. Uncertainty sources are related either to observational constraints or to the physical inputs (microphysics) of the models. Among observational constraints, uncertainties in element abundances, in particular from CNO elements, are the major source of uncertainty and dominate the error budget in model uncertainties of  helioseismic diagnostics. Reducing the uncertainty in spectroscopic measurements does not seem likely in the near future. Systematic uncertainties, hinted at by differences between AGSS09 and C11 values (Figure~\ref{fig:composition}), are a reason for concern. These are related to the choice of spectral lines used by the different authors, the underlying solar atmosphere, and methods for spectroscopic analysis. See for example \cite{bergemann:2021} for an extensive discussion of the solar oxygen abundance determination, which is dominated by systematic uncertainties (see also \cite{cubasarmas:2020}).

Regarding microphysics, radiative opacities remain the most uncertain and critical for solar models. The simultaneously high temperature and density in the solar interior are still not reachable by experiments in a systematic way. Therefore, atomic radiative opacities for solar models rely completely on theoretical calculations. OPAL \cite{opal:1996} and OP \cite{badnell:2005} are widely used sources. The differences between the two have been used as a measure of the uncertainty \cite{villante:2014}, yielding values from 5\% at the base of the convective envelope to 2\% in the solar core. The solar abundance problem, however, requires an increase of at least 15\% of the opacity at the base of the convective zone from either OPAL or OP  to make low-Z SSMs consistent with helioseismic results \cite{bahcall:2004,jcd:2009, guzik:2010, villante:2010, villante:2014, song:2018}. New atomic opacities (OPLIB) have been presented by the Los Alamos group \cite{colgan:2016} and by the OPAS collaboration \cite{blancard:2012, mondet:2015}, but differences are similar to those between OP and OPAL, except that OPLIB opacities seem too small in the solar core, leading to lower core temperatures and $^8$B and $^7$Be neutrinos fluxes too low to be compatible with experimental results. 

The first measurement of radiative opacities at conditions similar to those at the base of the solar convective zone was carried out for iron with the Z machine at Sandia National Laboratories and reported in \cite{bailey:2015}. Experimental results yielded larger  opacity, by about 40\% on the Fe contribution to opacity, than any theoretical calculation. The differences are dominated by a large systematic discrepancy in the quasicontinuum opacity. This result alone implies an increase of 7\% in the total opacity of a solar mixture at the base of the convective zone, where Fe is a main contributor \cite{basu:2008}. The origin of the discrepancy between models and experiment is not yet understood. Further experiments for Cr and Ni were carried out by the same group \cite{nagayama:2019} to enhance understanding. Results point towards problems in theoretical calculations related to, among others, the treatment of line broadening \cite{krief:2016a,krief:2016b,ovechkin:2019} and atom-plasma interactions. However, some experimental results remain not understood, such as the behaviour of the quasicontinuum as a function of the atomic charge of the nucleus in consideration. Overall, the situation regarding radiative opacities is puzzling, and future experimental work is urgently needed \cite{hoarty:2019,perry:2020}, as well as further development in theoretical calculations \cite{nahar:2016,krief:2016b,more:2017,krief:2018,baggott:2020}.

Other uncertainty sources related to microphysics are better understood or have a smaller impact on solar model predictions. Uncertainties in the equation of state, however, might have an impact on helioseismic determinations of solar abundances \cite{vorontsov:2013,buldgen:2017,jcd:2020}. Uncertainty in the gravitational settling rates of heavy elements and helium are estimated to be about 15-20\% \cite{thoul:1994}, although this is a rough estimate and relies to some extent on phenomenological constraints such as the surface helium abundance in the Sun. Uncertainties in nuclear reaction rates are discussed in more detail in Sec.~\ref{sec:Nuclear}.

Uncertainty sources can be grouped according to the way in which they affect solar neutrino predictions: nuclear reactions affecting individual neutrino fluxes; environmental factors affecting the solar core temperature; and the abundance of C, N, and O, which directly affects the fluxes in the CNO cycle (see \cite{haxton:2008,serenelli:2013,serenelli:2016} for details). The last three columns in Table~\ref{tab:solarnus} list the total uncertainty for each neutrino flux from each of these three classes, with results taken from \cite{vinyoles:2017}. Note that metals other than CNO are included as environmental uncertainties, as they affect the model neutrino predictions only through their contribution to the radiative opacity in the solar core. 
\begin{marginnote}[]
\entry{The composition and opacity}{represent the main general uncertainty sources for solar models.}
\entry{Nuclear cross section uncertainties}{matter for specific neutrino fluxes.}
\entry{Opacity calculations}{disagree with the only available opacity measurement carried out so far under solar conditions. }
\end{marginnote}

%% file: nuclear.tex
\section{Nuclear reactions in the Sun \label{sec:Nuclear}} 

In this section, the state of the art on nuclear reactions in the Sun is reviewed, and recommendations are developed. This text generally follows the approach of the decadal ``Solar Fusion Cross Sections'' community meeting based reviews, here called SFI and SFII. SFI and SFII include original works until 1997 \cite{1998RvMP...70.1265A} and 2009 \cite{2011RvMP...83..195A}, respectively. A third edition is planned for 2022. 

\subsection{Reactions important for solar neutrinos}
\label{subsec:Nuclear:Intro}

At the temperature of the solar core, only hydrogen burning is relevant, and the proton-proton chains (pp chains) dominate  \cite{2006ApJS..165..400B}. For the description of the nuclear reactions inside these chains, the following shorthand notation is adopted here:
\begin{equation}
^3{\rm He}(\alpha,\gamma)^7{\rm Be} \qquad \equiv \qquad ^3{\rm He} + \alpha \longrightarrow \gamma  + ^7{\rm Be}
\end{equation}
Here, $\beta^+$, electron capture, and $\alpha$ decays are denoted as $(e^+ \nu_e)$, $(e^-,\nu_e)$, and $(\alpha)$.

The three pp chains, called pp-I, pp-II, and pp-III, respectively, dominate energy production (Figure \ref{fig:Nuclear:pp}, section \ref{subsec:Nuclear:pp}). 
The second process of hydrogen burning, the carbon-nitrogen-oxygen (CNO) cycle, consists of the CN cycle and the NO cycle, and produces the so-called CNO neutrino fluxes (Figure \ref{fig:Nuclear:CNO}, section \ref{subsec:Nuclear:CNO}). 

For all the nuclear reactions considered here, the Coulomb barrier given by electrostatic repulsion between the two positively charged reaction partners far exceeds the kinetic energy of the thermal motion of the reaction partners in the solar core, even considering the high-energy tails of their thermal Maxwell-Boltzmann distribution. Below the Coulomb barrier, the dependence of the nuclear reaction cross section $\sigma(E)$ on center-of-mass energy $E$ can be parameterized using the so-called astrophysical S-factor $S(E)$ \cite{1957RvMP...29..547B}:
\begin{equation}\label{eq:Nuclear:Sfactor}
\sigma(E) = \frac{1}{E} S(E) \exp \left[  -\frac{b}{\sqrt{E}} \right]
\end{equation}
with $b=-2\pi Z_1 Z_2 \alpha \sqrt{\mu c^2 / 2}$ for particles with nuclear charges $Z_{1,2}$, masses  $m_{1,2}$, and reduced mass $\mu=m_1m_2/(m_1+m_2)$, $\alpha$ the fine structure constant and $c$ the vacuum speed of light. $S(E)$ varies only weakly with energy and encodes the strictly nuclear parts of the cross section. The thermonuclear reaction rate $N_{\rm A} \langle \sigma v \rangle$ is then given by the product of the S-factor (\ref{eq:Nuclear:Sfactor}) and the Maxwell-Boltzmann distribution for the temperature $T$:
\begin{equation} \label{eq:Nuclear:TNRR}
N_{\rm A} \langle \sigma v \rangle = N_{\rm A} \underbrace{\sqrt{\frac{8}{\mu\pi}}(k_{\rm B}T)^\frac{3}{2}}_{\rm Maxwell} S(E) \int\limits_{0}^{\infty} \exp \left[  \underbrace{-\frac{b}{\sqrt{E}}}_{\rm Coulomb}  \underbrace{-\frac{E}{k_{\rm B}T}}_{\rm Maxwell} \right] dE
\end{equation}
In equation (\ref{eq:Nuclear:TNRR}), the energy-dependent factors are labeled with their origins from the Maxwell-Boltzmann distribution or the Coulomb barrier.

\begin{marginnote}[]
\entry{Astrophysical S-factor $S(E)$}{Low-energy parameterization of the energy-dependent cross section $\sigma(E)$, given by equation (\ref{eq:Nuclear:Sfactor}).}
\entry{Thermonuclear reaction rate}{Number of nuclear reactions per time and volume, given by equation (\ref{eq:Nuclear:TNRR}).}
\end{marginnote}

The maximum of the integrand of $N_{\rm A} \langle \sigma v \rangle$ is called the Gamow peak and lies at $E$ = 6-28~keV, depending on the precise reaction. It is always above the central solar temperature of 1.4 keV but far below the respective Coulomb barrier of 400-2100 keV.
As a result, for most nuclear reactions, the cross section is so low that there are no experimental data directly at the energies relevant for solar fusion, i.e. at the Gamow peak. A notable exception is the $^2$H($p,\gamma$)$^3$He reaction \cite{2002NuPhA.706..203C, 2020Natur.587..210M}. For all other nuclear reactions, experimental data must be taken at the lowest possible energies, including at underground accelerators \cite{2010ARNPS..60...53B, 2018PrPNP..98...55B,Bemmerer18-SNC}, and then extrapolated down to the solar Gamow peak energy.  

One possible approach to such extrapolations are R-matrix fits \cite{1958RvMP...30..257L, 2010RPPh...73c6301D}. There, experimental data from many reaction channels are described in a consistent framework and then extrapolated.
This has been attempted, for example, for the $^{14}$N($p,\gamma$)$^{15}$O reaction \cite{2011RvMP...83..195A}.
An alternative approach to derive the low-energy cross section is given by so-called {\it ab initio} calculations, which have been reported for pp-chain reactions such as $^2$H($p,\gamma$)$^3$He \cite{2016PhRvL.116j2501M} and $^3$He($\alpha,\gamma$)$^7$Be \cite{2011PhRvL.106d2502N, 2016PhLB..757..430D, 2019PhRvC.100b4304V}. 

At the low energies relevant for solar fusion, electron screening, or electron shielding, reduces the repulsive electric potential of the target nucleus. This effect is different for electrically neutral atoms in the laboratory \cite{1987ZPhyA.327..461A} from the plasma in the center of the Sun \cite{1954AuJPh...7..373S}. 
Unexpectedly high laboratory electron screening has been reported for some light-ion reactions \cite{2004EPJA...19..283R,2008PhRvC..78a5803H,2015PhRvC..92f5801C}. These effects are not strong enough to significantly change the solar fusion reaction cross sections \cite{1998RvMP...70.1265A, 2011RvMP...83..195A, 2010ARNPS..60...53B, 2018PrPNP..98...55B}, but they are in tension with the general screening framework \cite{1954AuJPh...7..373S}. Experiments at high-power lasers \cite{2020PhRvC.101d2802Z}, which are essentially screening-free, seem to confirm stellar extrapolations of classical ion beam experiments with standard screening corrections \cite{1954AuJPh...7..373S}.
In principle, plasma effects may also affect the rate of nuclear decays, but are not expected to lead to large deviations \cite{1997ApJ...490..437G}. 

\subsection{Nuclear reactions affecting the pp-chain solar neutrinos}
\label{subsec:Nuclear:pp}

The rate of all three pp chains, hence overall energy production and the equilibrium temperature of the Sun, is controlled by the initial reaction, $^1$H($p,e^+\nu_e$)$^2$H. 
Its cross section is many orders of magnitude too low to be accessible experimentally. However, theoretical work has converged to an accepted value with an uncertainty as low as 1\% \cite{2011RvMP...83..195A}, in agreement with more recent calculations on the lattice \cite{2017PhRvL.119f2002S}. The subsequent reaction, $^2$H($p,\gamma$)$^3$He, proceeds much faster, based on highly precise underground data \cite{2002NuPhA.706..203C,2020Natur.587..210M}, and thus does not limit the pp chains.

The intersection between the pp-I and pp-II chains is given by the competition between the $^3$He($^3$He,2p)$^4$He (pp-I) and $^3$He($\alpha,\gamma$)$^7$Be (pp-II) reactions, which occur in about 83\% and 17\% of the cases, respectively. For the former of these reactions, a LNGS-based cross section measurement at LUNA (Laboratory for Underground Nuclear Astrophysics) ruled out a previously postulated resonance \cite{1999PhRvL..82.5205B}. 
The latter reaction has been studied over a wide energy range, but not yet at solar energies \cite[and references therein]{1959PhRv..113.1556H,2006PhRvL..97l2502B, 2013PhRvC..87a5802D,2019PhRvC..99e5804S}. Taking into account uncertainties from the extrapolation, its solar rate is believed to be known with 5\% uncertainty, using the weighted average of all the experiments \cite{2011RvMP...83..195A}. A further improvement hinges on theoretical \cite{2020JPhG...47e4002Z} and experimental work connecting the well-studied 1-MeV interaction energy range to the solar Gamow peak at $\sim$0.02 MeV.

\begin{figure}
    \begin{tabular}{p{0.5\textwidth}p{0.5\textwidth}}
    \includegraphics[width=0.5\textwidth]{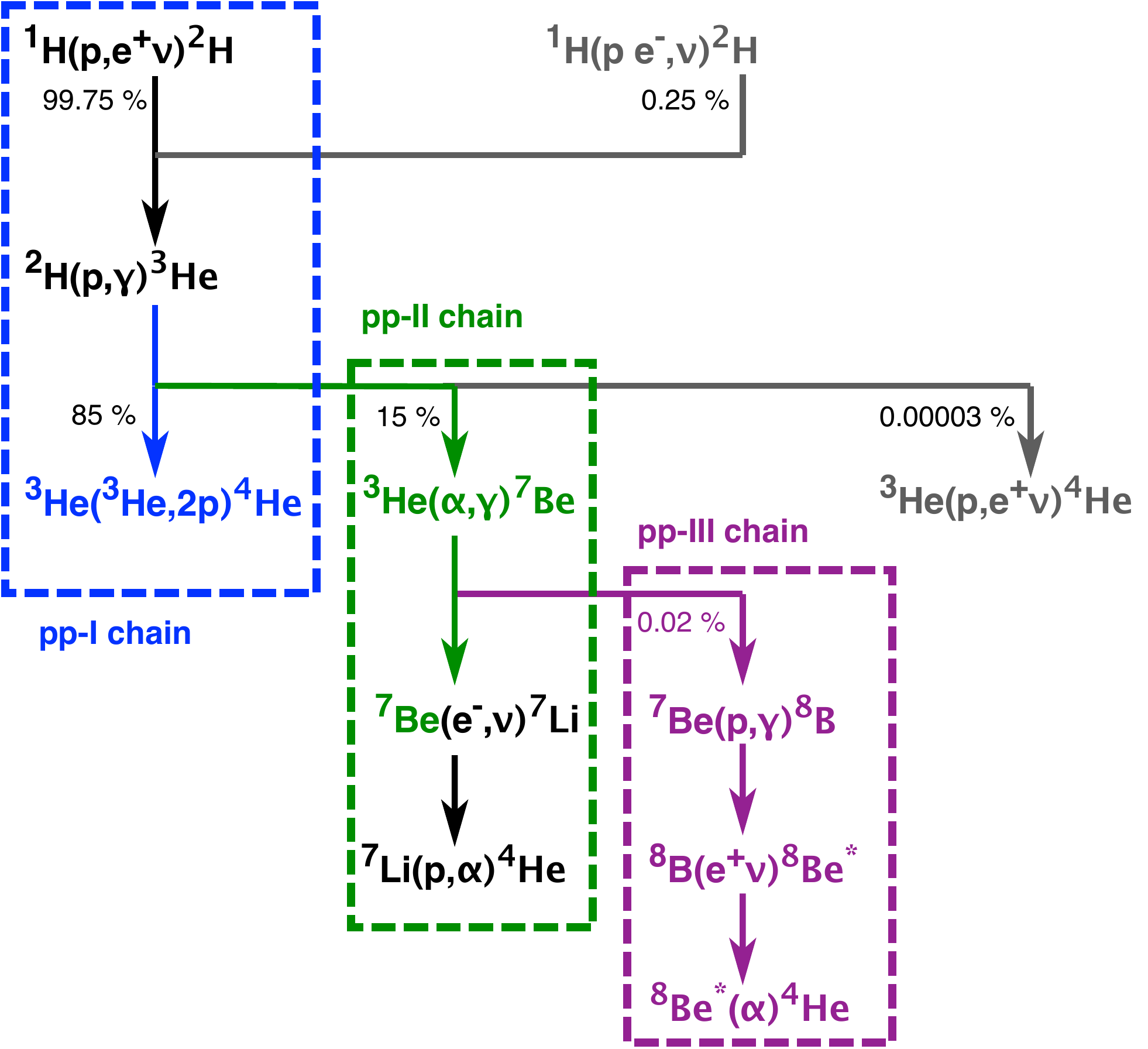}%
    &
    \includegraphics[width=0.5\textwidth,trim=5mm 0 0 0,clip]{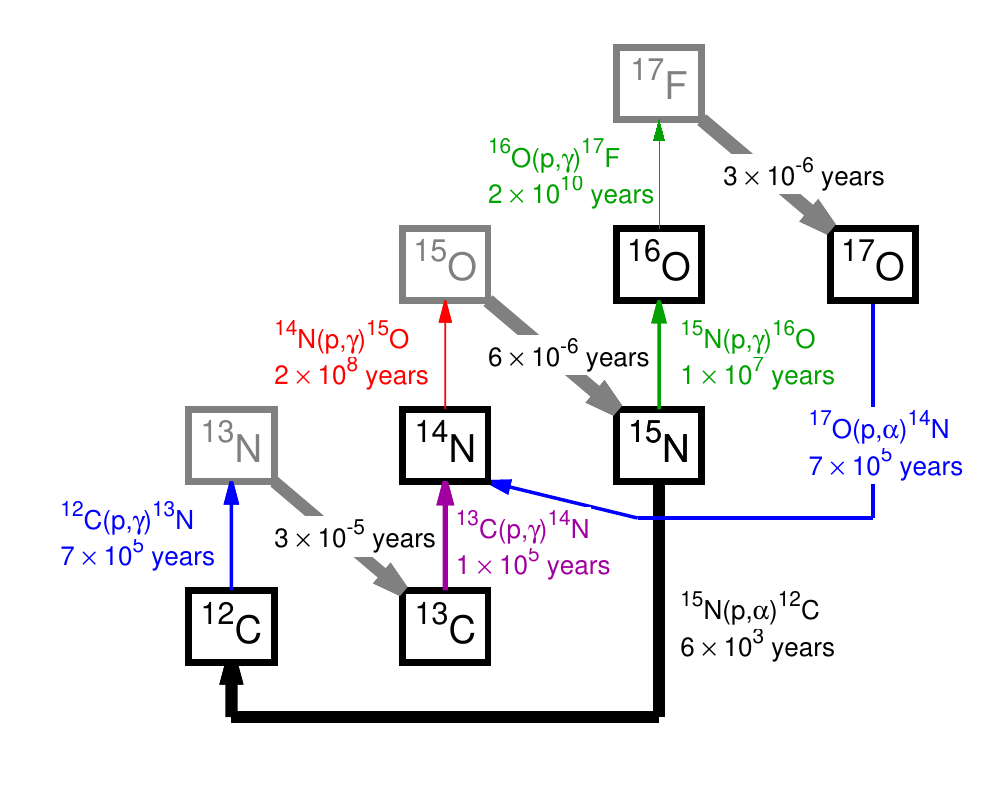}
    \\
    \end{tabular}
    \caption{Nuclear reactions in the Sun. (Left) Proton-proton chains.
    (Right) CNO cycle. Here, effective lifetimes of the starting nuclide against this nuclear reaction ($\tau_{\rm reaction}=1/\rho X_{\rm H}N_{\rm A}\langle\sigma v\rangle$) or  decay ($\tau_{\rm decay}=1/\lambda$) are given. $\rho$ is the solar core density, $X_{\rm H}$ the hydrogen mass fraction. 
    and $\lambda$ is the decay constant. Wider arrows represent faster transmutations.
    }
    \label{fig:Nuclear:pp}
    \label{fig:Nuclear:CNO}
\end{figure}

\begin{marginnote}
\entry{Overall rate of pp-I, -II, and -III}{The overall rate of all the pp chains is given by the $^1$H($p,e^+\nu_e$)$^2$H reaction and on solid theoretical ground, with 1\% uncertainty.} 
\entry{Intersection between pp-I and pp-II}{The intersection between the pp-I and pp-II chains is controlled by the $^3$He($\alpha,\gamma$)$^7$Be reaction, with 5\% uncertainty.}
\entry{Intersection between pp-II and pp-III}{The ratio of the pp-II and pp-III chains depends on the $^7$Be($p,\gamma$)$^8$B reaction, with 8\% uncertainty.}
\end{marginnote}

The rate of the third branch, pp-III, is much lower, 0.00002\%, and given by the competition between the electron capture decay of $^7$Be (pp-II) and the $^7$Be($p,\gamma$)$^8$B reaction (pp-III). 
Due to its low cross section and the presence of a strong resonance that complicates the extrapolation, the $^7$Be($p,\gamma$)$^8$B reaction is difficult to study in the laboratory. Its accepted rate is 8\% precise \cite{2011RvMP...83..195A} and mainly based on an experiment using a proton beam incident on radioactive $^7$Be targets \cite{2002PhRvL..88d1101J,2010PhRvC..81a2801J}. However, radioactive $^7$Be beam data hints at a lower cross section \cite{2018EPJA...54...92B}, so further work may be needed.

Finally, two additional nuclear reactions branch out to and from the three main pp chains in Figure \ref{fig:Nuclear:pp}: $^1{\rm H}(p e^-,\nu_e)^2{\rm H}$ (pep) is an alternative starting point for all three chains; and $^3$He($p,e^+ \nu_e$)$^4$He (hep) is an alternative termination to the pp-I chain. Both give rise to low neutrino fluxes, and neither can be studied in the laboratory.

\subsection{Nuclear reactions affecting the CNO solar neutrinos}
\label{subsec:Nuclear:CNO}

The CNO cycle \cite{Weizsaecker38-PZ,1939PhRv...55..103B} (right panel, figure \ref{fig:Nuclear:pp}) starts from pre-existing $^{12}$C in the solar core. 
In equilibrium, the lifetime of the cycle is dominated by its slowest reaction, $^{14}$N($p,\gamma$)$^{15}$O. This reaction takes more than 99\% of the integrated time of all the six transmutations in the cycle, so that almost all the initial $^{12}$C is transmuted to $^{14}$N and stored there.

Two other nuclear reactions also play interesting roles in CNO burning: $^{12}$C($p,\gamma$)$^{13}$N and $^{16}$O($p,\gamma$)$^{17}$F. The temperature dependence of the former is less steep than for the $^{14}$N($p,\gamma$)$^{15}$O case. As a result, in the early Sun and also in the outer layers of the present-day solar core, this reaction controls the onset of CNO burning, causing a double-peaked structure in the radial emission profile of the $^{13}$N neutrinos \cite{2006ApJS..165..400B}, as observed in Figure~\ref{fig:nuprofiles} in Sec.~\ref{s:models2}. 

The CN- and NO-cycles intersect at $^{15}$N, but the 2000 times higher rate of $^{15}$N($p,\alpha$)$^{12}$C compared to $^{15}$N($p,\gamma$)$^{16}$O \cite{2009JPhG...36d5202B, 2010PhRvC..82e5804L} hinders the passage of nucleosynthetic material between these cycles in the Sun. Instead, pre-existing $^{16}$O feeds the $^{16}$O($p,\gamma$)$^{17}$F reaction in the Sun. 
The flux of $^{17}$F neutrinos therefore depends on the initial $^{16}$O abundance of the Sun and the (very slow) $^{16}$O($p,\gamma$)$^{17}$F reaction. Any $^{17}$F produced is quickly returned to the $^{14}$N reservoir by way of the $^{17}$O($p,\alpha$)$^{14}$N reaction \cite{2016PhRvL.117n2502B}. 

\begin{marginnote}
\entry{CN cycle}{The CN cycle begins and ends at $^{12}$C, passing through $^{14,15}$N.}
\entry{NO cycle}{The NO cycle begins at $^{15}$N and returns to $^{14}$N. In the Sun, it mainly starts from pre-existing $^{16}$O.}
\entry{CNO cycle}{The ensemble of the CN- and NO-cycles. Together, they contribute only 0.8\% to the solar luminosity.}
\end{marginnote}

Due to its paramount importance for the rate of the CN cycle and, hence, the predicted integral flux of $^{13}$N and $^{15}$O neutrinos, the $^{14}$N($p,\gamma$)$^{15}$O reaction has been studied many times \cite[and references therein]{2016PhRvC..93e5806L,2018PhRvC..97a5801W} since it was initially proposed by Bethe \cite{1939PhRv...55..103B} and Weizsäcker \cite{Weizsaecker38-PZ}. The  $^{14}$N($p,\gamma$)$^{15}$O rate was reduced by a factor of two when comparing SFI and SFII \cite{1998RvMP...70.1265A,2011RvMP...83..195A}. This strong revision was due to an even stronger reduction in the contribution by capture to the ground state of $^{15}$O, based on indirect experiment, theory, and direct experiment \cite{2001PhRvL..87o2501B,2001NuPhA.690..755A,2004PhLB..591...61F,2005EPJA...25..455I,2005PhRvL..94h2503R,2008PhRvC..78b2802M,2011PhRvC..83d5804M}. 

The latest community-based extrapolated zero-energy S-factor is $S_{1,14}(0) = (1.60\pm0.09)$~keV~barn, using the SFII R-matrix analysis \cite{2011RvMP...83..195A}, here corrected for an updated strength of the normalization resonance \cite{2016PhRvC..94b5803D}. Of the two most recent individual studies, one 
hints at a somewhat higher S-factor \cite{2016PhRvC..93e5806L}. As a result, here $S_{1,14}(0)$ = (1.60$\pm$0.13) keV~barn is recommended, i.e. an error of 8\%, so that the recent results are included in 2$\sigma$.

\begin{figure}
    \centering
    \includegraphics[width=0.8\textwidth]{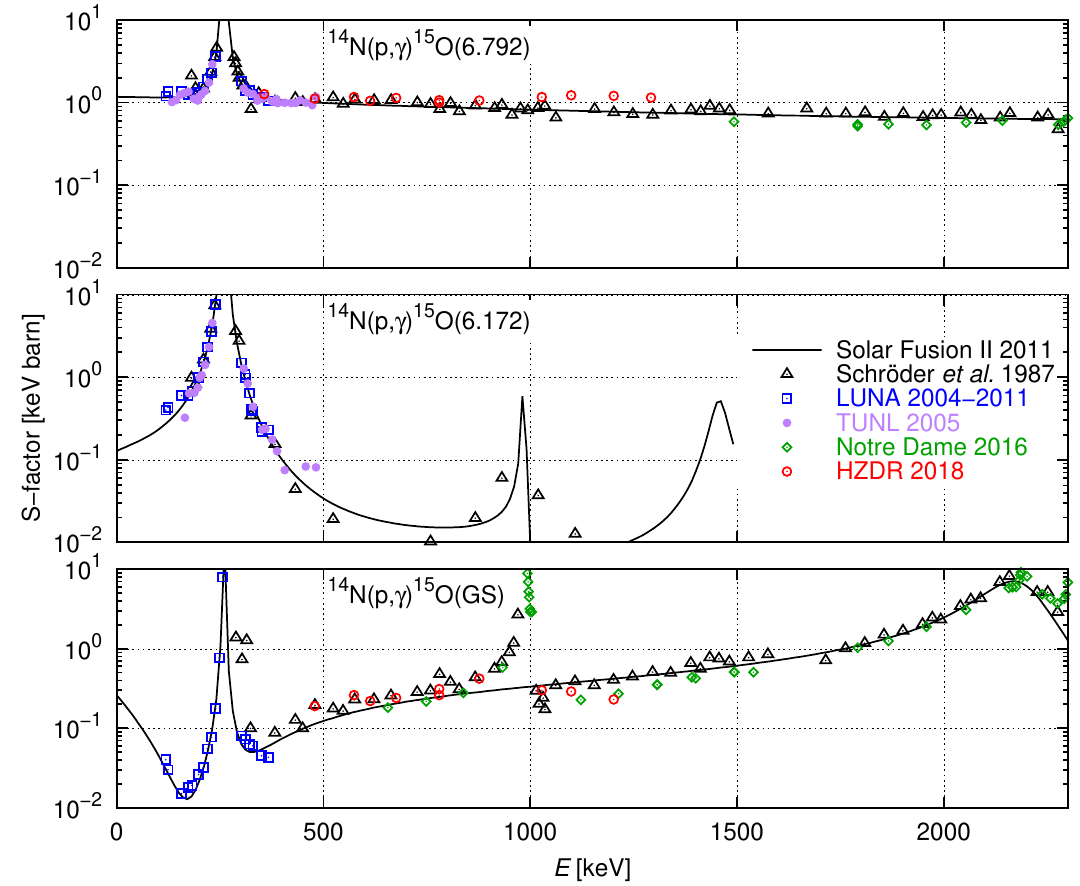}
    \caption{Astrophysical S-factor of the $^{14}$N($p,\gamma$)$^{15}$O reaction as a function of center-of-mass energy $E$. The three most important transitions are shown: capture to the $^{15}$O excited states at 6.792 and 6.172 MeV and to the ground state (top to bottom). Experimental data is by Schröder {\it et al.} \cite{1987NuPhA.467..240S}, LUNA \cite{2004PhLB..591...61F,2005EPJA...25..455I,2008PhRvC..78b2802M,2011PhRvC..83d5804M}, TUNL \cite{2005PhRvL..94h2503R}, Notre Dame \cite{2016PhRvC..93e5806L}, and HZDR \cite{2018PhRvC..97a5801W}. The lines are R-matrix fits by SFII \cite{2011RvMP...83..195A} and precede the latest two experiments \cite{2016PhRvC..93e5806L,2018PhRvC..97a5801W}.}
    \label{fig:Nuclear:S114}
\end{figure}

The rate of the $^{12}$C($p,\gamma$)$^{13}$N reaction is more uncertain, at 16\% \cite{1999NuPhA.656....3A,2011RvMP...83..195A}. The last comprehensive study of this reaction dates back to 1974 \cite{1974NuPhA.227..291R}. 

For  $^{16}$O($p,\gamma$)$^{17}$F, there is an evaluated S-factor curve with 6-7\% uncertainty at high energy -- 0.5-2.5 MeV in the laboratory \cite{2012NIMPA.688...62M}. At the energies relevant here (the Gamow peak energy), Solar Fusion II recommends a slightly higher error, 8\% \cite{2011RvMP...83..195A}, based on theory and extrapolations.

Unlike the pp-chain reactions, no {\it ab initio} theoretical description has yet been reported for any of the CNO nuclear reactions, even though this mass range has recently become at least in principle accessible \cite{2014PhRvL.112j2501E}.

\subsection{Recommended future work}
\label{subsec:Nuclear:Recommend}

New experiments and new theoretical work should go hand in hand to improve the precisions of the rates of the pp-chain reactions $^3$He($\alpha,\gamma$)$^7$Be and $^7$Be($p,\gamma$)$^8$B to 3\%, so as to match the recent $^2$H($p,\gamma$)$^3$He case \cite{2002NuPhA.706..203C,2020Natur.587..210M}.
For the three key CNO reactions $^{14}$N($p,\gamma$)$^{15}$O, $^{12}$C($p,\gamma$)$^{13}$N, and $^{16}$O($p,\gamma$)$^{17}$F, new data are needed to bring the precision to 5\%. In addition, new theoretical approaches should be extended to address these cases.
Finally, new capabilities offered by high-power lasers should be used to study the radiative opacities of C, N, and O, and also higher charge number atoms, in the laboratory, as well as plasma effects on nuclear reactions and decays.

%% file: models2.tex
\section{SOLAR NEUTRINO FLUXES \label{s:models2}}

The theoretical solar neutrino spectrum in Figure~\ref{fig:nuspectrum} shows the five fluxes associated with the  pp chain, the continuum fluxes from the $\beta^-$ decay of $^{13}$N, $^{15}$O, and $^{17}$F in the CNO cycle, and those from the  mono-energetic $e^-$-captures on the same isotopes \cite{bahcall:1990, stonehill:2004, villante:2015}. 
In this review, we denote neutrino fluxes at the Earth as $\Phi(X)$, where $X$ will denote the specific neutrino flux.

Global analyses of solar and terrestrial  experimental neutrino data have been used to determine the solar neutrino fluxes \cite{fogli:2004,bahcall:2004b}. In the last decade, Borexino has played a fundamental role after publishing initial results for the $^7$Be flux~\cite{arpesella:2008}, as nicely illustrated first by \cite{gonzalez-garcia:2010}. This work has been updated by \cite{bergstrom:2016} by including all experimental neutrino data available until 2016. These experimental solar neutrino fluxes are reported as ``no LC'' (no luminosity constraint) in Table~\ref{tab:solarnus}.

The energy produced by nuclear reactions in the Sun can be determined from the neutrino fluxes:
\begin{equation}
\frac{L_{\rm nuc}}{(1 {\rm A. U.})^2} = \sum_{i=1,8} \alpha_i \Phi(X_i), 
\end{equation}
where the sum extends over all neutrino fluxes (neglecting ecCNO fluxes, see Fig.~\ref{fig:nuspectrum}), $\alpha_i$ represents the energy contribution of the reactions associated to each of the fluxes \cite{bahcall:2002,bergstrom:2016}, and A.U. is the astronomical unit. Replacing $\Phi(X_i)$ with experimental results, the nuclear energy production in the Sun is:
\begin{equation}
L_{\rm nuc} = 1.04^{+0.07}_{-0.08}\  \lsun. \label{eq:solarlumi}
\end{equation}
The uncertainty is dominated by the uncertainty of $\Phi({\rm pp})$. The latter is primarily determined by the contribution of all the gallium experiments to the global analysis and the constraining power of Borexino on $\Phi(^7{\rm Be})$.

Equation~\ref{eq:solarlumi} represents the most accurate and precise experimental determination of the origin of energy in the Sun, a quest that started more than a century ago. The need to improve this further stems from the possibility that non-standard channels might also be present. This is, for example, the case for ALPs (Sect.~\ref{subsec:unknown}). For some of these particles, e.g. axions, the most stringent upper limits on  energy losses from the Sun come from helioscopes. But in other cases, e.g. dark photons or millicharged particles, limits from solar models offer the most constraining power in regions of parameter space \cite{vinyoles:2015,vinyoles:2016}. These limits arise from a combination of solar neutrinos and helioseismic probes, and establish a maximum energy loss through the non-standard channels of 1 to 2\% of $\lsun$ to 1$\sigma$, much better than the current purely experimental result expressed in Eq.~\ref{eq:solarlumi}. But they are model dependent and,  to some extent, subject to uncertainties in the accuracy of solar models. A large improvement in the experimental result is highly desirable. For this, a precise measurement of $\Phi(\hbox{pp})$ is needed. 
If  experimental data are complemented by the solar luminosity constraint (LC) \cite{bahcall:2002, vescovi:2021}, which assumes that the solar luminosity is produced by nuclear reactions, the result is 
\begin{equation}
L_{\rm nuc} = 0.991^{+0.005}_{-0.005} + 0.009^{+0.004}_{-0.005}\ \lsun, 
\end{equation}
where the first term refers now to the energy originating from the pp-chains and the second one to energy from the CNO cycle. The individual fluxes resulting from this analysis are listed in Table~\ref{tab:solarnus} in the 'LC' column. The largest impact of including the LC occurs for $\Phi(\hbox{pp})$, which controls most of the energy production in the Sun, and  $\Phi(\hbox{pep})$, which is directly linked to $\Phi(\hbox{pp})$~\cite{bahcall:1969,2011RvMP...83..195A}.

\begin{table}
\begin{tabular}{cccccccrrr}
Flux & \multicolumn{2}{c}{Solar (Global)} &&  \multicolumn{2}{c}{SSM - B16} && \multicolumn{3}{c}{Uncertainties} \\ 
\cline{2-3} \cline{5-6} \cline{8-10} 
& (no LC) &  (LC) && high-Z & low-Z && Nucl. & Envir. &  CNO \\ \hline
$\Phi(\hbox{pp}) $  &6.21$\pm0.50$&  $5.971^{+0.037}_{-0.033}$&& 5.98(0.6\%)& 6.03(0.5\%) && 0.4\% & 0.4\% & 0.1\%  \\
$\Phi(\hbox{pep}) $  & 1.51$\pm0.12$ & $1.448\pm0.013$ & & 1.44(1\%) & 1.46(1\%) && 0.6\% & 0.8\% & 0.3\%\\
$\Phi(\hbox{hep}) $ & $19^{+12}_{-9}$ & $19^{+12}_{-9}$ && 7.98(30\%) & 8.25(30\%) && 30\% & 1.3\% & 0.4\% \\
$\Phi(^7\hbox{Be}) $ & 4.85$\pm0.19$& $4.80^{+0.24}_{-0.22}$ && 4.93(6\%)& 4.50(6\%) && 5.0\% & 4.1\% & 0.8\% \\
$\Phi(^8\hbox{B}) $ & $5.16^{+0.13}_{-0.09}$ & $5.16^{+0.13}_{-0.09}$ && 5.46(12\%) & 4.50(12\%) && 7.6\% & 9.2\% & 1.9\% \\
$\Phi(^{13}\hbox{N}) $ & $\leq 13.7$ & $\leq 13.7$ && 2.78(15\%)&  2.04(14\%) && 6.2\% & 6.9\% & 12\% \\
$\Phi(^{15}\hbox{O}) $ & $\leq 2.8$ &  $\leq 2.8$ && 2.05(17\%) &  1.44(16\%) && 8.7\% & 8.4\% & 12\%\\
$\Phi(^{17}\hbox{F}) $ & $\leq 85$ & $\leq 85$ && 5.29(20\%) & 3.26(18\%) & &9.3\% & 9.0\%& 16\%\\ \hline
$\chi2$ & & & & 6.0 & 7.0 & \\
\hline
\end{tabular} 
\caption{Solar neutrino fluxes. Solar: experimental results with and without the inclusion of the luminosity constraint (LC). B16 high-Z and B16 low-Z: SSMs results and uncertainties based on GS98 and AGSS09 solar mixtures. Last three columns: contribution to model uncertainties from different types of sources (see text for more information). Fluxes given in: $10^{10}$ (pp), $10^9$ ($^7$Be), $10^8$ (pep, $^{13}$N, $^{15}$O), $10^6$ ($^8$B, $^{17}$F), and  $10^3$ (hep) $\hbox{cm}^{-2}\hbox{s}^{-1}$. \label{tab:solarnus}}
\end{table}

Borexino has subsequently reported measurements of all the individual reactions in the pp chain \cite{2014Natur.512..383B, borexino:pp2, borexino:pp3}, including a direct measurement of $\Phi(\hbox{pp})=6.10\pm0.5^{+0.3}_{-0.5}\times10^{10}\hbox{cm}^{-2} \hbox{s}^{-1}$,  an upper limit $\Phi(\hbox{hep}) < 2.2\times10^{5}\hbox{cm}^{-2} \hbox{s}^{-1}$  flux, and the most stringent measurement of the $^7$Be flux, $\Phi(^7\hbox{Be})= 4.99\pm0.11^{+0.6}_{-0.8}\times10^{9}\hbox{cm}^{-2} \hbox{s}^{-1}$, where quoted uncertainties are statistical and systematic, respectively. Finally, Borexino has also provided the first ever direct measurement of the combined neutrino flux from the CNO cycles \cite{2020Natur.587..577B},  $\Phi(\hbox{CNO}) = \Phi(^{13}\hbox{N}) + \Phi(^{15}\hbox{O}) = 7^{+3}_{-2}\times10^{8}\hbox{cm}^{-2} \hbox{s}^{-1}$. 
A comparison with results of the global analysis without the LC (Table~\ref{tab:solarnus}) shows that  Borexino improves on some solar neutrino fluxes. However, a global analysis aimed at the determination of the solar neutrino fluxes using all neutrino data posterior to 2016 is missing in the literature, with recent work focusing on determination of the neutrino oscillation parameters \cite{esteban:2017,2020JHEP...09..178E}. 

\begin{figure}
\includegraphics[width=0.8\textwidth]{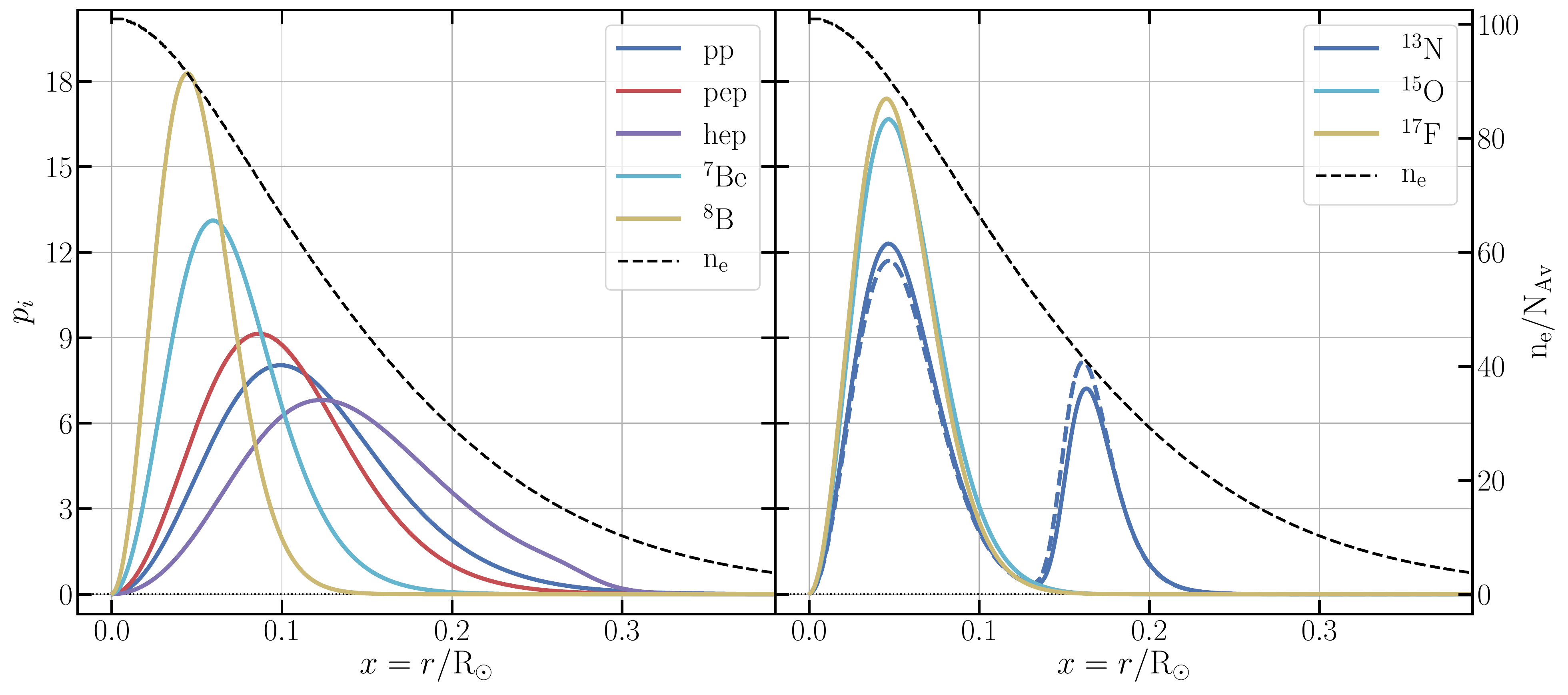}
\caption{Probability distribution function of production for all solar neutrino fluxes ($\int_0^1{p_i dx}=1$, with $i$ running over all fluxes). Left panel: neutrinos from the pp chain. Right panel: neutrinos from the CNO cycle. For $\Phi(^{13}\hbox{N})$ distributions from a high-Z (solid line) and a low-Z SSM (dashed line) are shown. Both panels show the electron density ($n_e$) distribution in units of cm$^{-3}$~mol$^{-1}$. \label{fig:nuprofiles}}
\end{figure}

SSM neutrino fluxes and uncertainties are also given in Table~\ref{tab:solarnus}, for the B16 high-Z and B16 low-Z models. Metals affect solar neutrinos by modifying the core temperature in the Sun and, as a result, the stronger the temperature dependence of neutrino fluxes, the larger the difference between the high-Z and low-Z predictions. Such dependence is primarily responsible for the distribution of the production of neutrino fluxes in the solar interior, as illustrated by the production probability distribution functions in Figure~\ref{fig:nuprofiles}. CNO fluxes carry an additional dependence on the C, N, and O abundance in the solar core. The distribution profiles are insensitive to the solar composition, with the  exception of $\Phi(^{13}\hbox{N})$. The external peak in its distribution is produced by the production of N from  primordial C in the Sun. It is therefore proportional to the C abundance, and largely independent of environmental factors and uncertainty in the $^{12}$C(p,$\gamma$)$^{13}$N reaction \cite{villante:2021}. The second, inner peak comes from the CN cycle operating in steady state, which depends not only on the total abundance of C+N, but also on the temperature of the solar core, i.e. it is affected by environmental factors. Its importance is therefore smaller in low-Z solar models, as seen in Figure~\ref{fig:nuprofiles}. Overall, the difference in radial distributions between the two models are very small and should have a negligible contribution to the integrated survival probability of $\Phi({\rm ^{13}N})$ neutrinos. The electron density, on which MSW effects depend, is also shown in both panels. 
\begin{marginnote}[]
\entry{Environmental factor}{Quantity that affects solar neutrino fluxes through its impact on the thermal structure of the solar core.}
\end{marginnote}

The possibility of using solar neutrinos to discriminate between low-Z and high-Z solar models with currently available solar neutrino data, limited to fluxes from the pp chain, leads to inconclusive results \cite{vinyoles:2017}, as given in the last row of Table~\ref{tab:solarnus}. Moreover, such comparison is sensitive only to the temperature in the solar core, i.e. to the radiative opacity, not directly the solar composition \cite{serenelli:2010}. Another possible test between low-Z and high-Z models, but also sensitive to core temperature, is the comparison of the ratio $R_{\rm I/II}$ \cite{borexino:pp2}, the relative intensity of the pp-I and pp-II chains, which is determined experimentally as  $R_{\rm I/II} = 2 \Phi(^7\hbox{Be})/\left(\Phi(\hbox{pp})-\Phi(^7\hbox{Be}) \right)$. Borexino results yield $R_{\rm I/II} = 0.178^{+0.027}_{-0.023}$ and the global fit $0.176\pm0.015$. Results for the SSMs are $0.180\pm0.012$ and $0.161\pm0.011$ for the B16 high-Z and B16 low-Z models respectively. Current experimental results seem to favor SSMs with higher core temperatures, as stated by \cite{borexino:pp2}.

\begin{marginnote}[]
\entry{Solar neutrino fluxes}{from the pp chain are sensitive to the thermal structure of the solar core. Fluxes from CNO cycle are directly sensitive to composition.}
\entry{Both high-Z and low-Z solar models}{agree with current experimental solar neutrino fluxes, with a slight preference for a hotter solar core, similar to that of high-Z SSMs.}
\entry{The nuclear origin of solar luminosity}{ is established  experimentally to 7\%.}
\entry{All pp-chain neutrino fluxes}{except hep, have now been measured.}
\entry{CNO-cycle neutrinos}{have been detected for the first time by Borexino.}
\end{marginnote}
Using solar neutrinos to break the degeneracy between opacities, i.e. solar core temperature, and composition, is only possible with neutrinos from the CNO cycle \cite{haxton:2008, serenelli:2010, serenelli:2013, gough:2019}. By separating the dependence on environmental factors, nuclear reactions and CN abundances, a relation between $\Phi(^8\hbox{B})$ and $\Phi(^{13}\hbox{N})$ or $\Phi(^{15}\hbox{O})$ (or a linear combination of the two) can be established, with the role of solar models limited to that of scaling factors \cite{haxton:2008, serenelli:2013}. Recently, \cite{agostini:2020,villante:2021} have determined this relation taking into account the differential sensitivity of Borexino to $^{13}\hbox{N}$ and $^{15}\hbox{O}$ neutrinos, obtaining
\begin{equation}
\frac{\Phi^{BX}(\hbox{CN})}{\Phi^{BX}_{\hbox{SSM}}(\hbox{CN})} = \left(\frac{\Phi(^8\hbox{B})}{\Phi_{\hbox{SSM}}(^8\hbox{B})}  \right)^{\alpha} \left(\frac{X_{\hbox{C}}}{X_{\hbox{C,SSM}}}\right)^{0.814}
\left(\frac{X_{\hbox{N}}}{X_{\hbox{N,SSM}}}\right)^{0.191} \left( \pm 9.1\% (\hbox{nucl}) \pm 0.5\% (\hbox{env}) \right), \label{eq:cnb8_1}
\end{equation}
where $\Phi^{BX}(\hbox{CN}) = (1-\xi) \Phi(^{13}\hbox{N}) + \xi \Phi(^{15}\hbox{O})$, with $\xi=0.764$, is the combination of fluxes that Borexino is sensitive to, $X_{\hbox{C}}$, $X_{\hbox{N}}$ denote carbon and nitrogen mass fractions, and the SSM subindex denotes SSM values. The relation can be simplified further to an almost linear relation between $\Phi^{BX}(\hbox{CN})$ and $(X_{\hbox{C}}+X_{\hbox{N}})$ if the fractional change of C and N with respect to the values in the SSM is assumed to be the same \cite{serenelli:2013}. Note that  dependencies on all environmental factors are almost perfectly cancelled out by the relation with $\Phi(^8\hbox{B})^{\alpha}$, which makes the above expression rather insensitive to uncertain quantities such as radiative opacities, and also valid  beyond the framework of SSMs. The value of the exponent, $\alpha=0.716$, is specific to $\Phi^{BX}(\hbox{CN})$. Analogous relations can be easily obtained \cite{serenelli:2013, villante:2021} for future detectors simply by determining $\xi$ and $\alpha$ according to the differential sensitivity of the detector to $\Phi(^{13}\hbox{N})$ and $\Phi(^{15}\hbox{O})$. If future experiments allow separate measurements of $\Phi(^{13}\hbox{N})$ and $\Phi(^{15}\hbox{O})$, the difference $\Phi(^{13}\hbox{N}) - \Phi(^{15}\hbox{O})$ can be used to determine the C abundance independently of N, which can then also be determined. 

Motivation for measuring neutrino fluxes from the CNO cycle goes well beyond the solar abundance problem. After a very short initial phase of about 1~Myr according to SSMs, and even shorter in other solar formation scenarios \cite{wuchterl:2001}, the solar core has been isolated from the rest of the solar system. Its composition is a fossil record of the primordial composition of the cloud from which not just the Sun, but all planets, formed. Moreover, the comparison between the core and surface abundance of metals could also be used to determine the efficiency of mixing processes in the Sun, processes for which there are no direct constraints so far, and that are needed for precision solar and stellar modeling.

%% file: solarphysics.tex
\section{SOLAR NEUTRINO PHYSICS }\label{s:physics}

\subsection{Solar neutrino flavour change}

\begin{figure}
    \centering \includegraphics[width=0.8\textwidth]{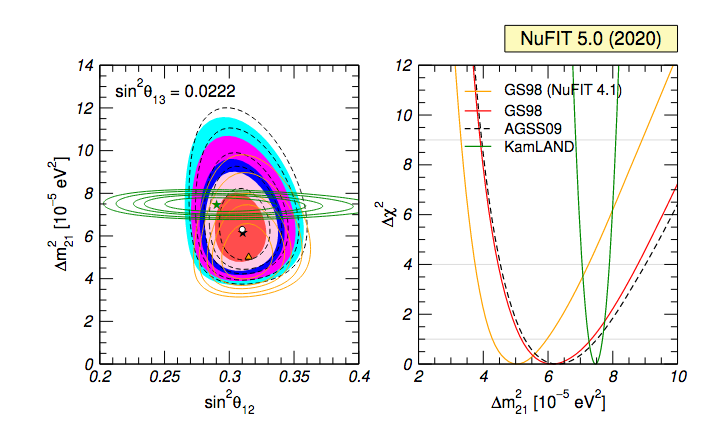}
    \caption{Current status of solar neutrino oscillation parameters.
    Left: Allowed parameter regions (1 $\sigma$ at 90 percent, 2$\sigma$ at 99 percent and 3 $\sigma$ CL for 2 degrees of freedom) from the combined analysis  of solar data for the GS98 model \cite{1998SSRv...85..161G} (full regions with best fit  marked by black star) and AGSS09 model \cite{2009ARA&A..47..481A} dashed void contours with best  fit marked by a white dot), and for the
analysis of KamLAND data (solid green contours with best fit marked by a green star) for fixed $\sin^2 \theta_{13} = 0.0222 (\theta_{13}=8.6^{\circ}$).
Also shown are as orange contours the previous results of the global analysis for the GS98 model \cite{1998SSRv...85..161G}.
Right: $\Delta \chi^2$ dependence on $\Delta m^2_{21}$ for the same four analyses after marginalizing over $\theta_{12}$ \cite{2020JHEP...09..178E}. Graph with kind permission of T. Schwetz.
    \label{fig:PMNS}}
\end{figure}

Observations of the Z$^0$ boson decay width show that there are only three neutrinos with masses less than half of the Z$^0$ width, i.e. 45 GeV. This is consistent with limits from Big Bang Nucleosynthesis (BBN)~\cite{Izotov_2010}.
Hence, flavour and mass eigenstates are linked with a unitary 3x3 matrix, called the Pontecorvo–Maki–Nakagawa–Sakata, or PMNS matrix \cite{1962PThPh..28..870M}.  Any 3x3 matrix can be parameterised by three mixing angles $\theta_{12},\theta_{13}$ and $\theta_{23}$ and a complex phase $e^{i\delta}$.  (A Majorana mass term for neutrinos would introduce two additional complex phases, but these are relevant only for neutrino-less double beta decay -- NLDBD -- searches.)
Hence, there are mass eigenstates ($\nu_1,\nu_2$) and ($\nu_3$) and flavour eigenstates ($\nu_e, \nu_\mu$ and $\nu_\tau$), which are linked by the PMNS matrix. This is 
analogous to the well known CKM matrix in the quark sector. The values of the matrix elements have to be determined experimentally. Results of mass eigenstates $m_{i,i=1,2,3}$ are presented in the form of $\Delta m^2_{ij} = m_j^2 - m_i^2$. Besides solar data, also reactor, astrophysical, atmospheric and long-baseline neutrino beams contribute to the determination of the PMNS matrix. The current status of its element values can be found in~\cite{2020JHEP...09..178E}, and seen graphically in Figure~\ref{fig:PMNS}.
Based on all available solar neutrino data, a consistent picture appears for the survival probability as mentioned in Section~\ref{s:intro}. 

Global solar neutrino data show that matter effects play a critical role. 
The weak scattering of $\nu_e$ off electrons in the Sun has a larger cross section than $\nu_{\mu/\tau}$, due to the presence of CC as well as NC channels.  This results in an additional effective mass term, which modifies the effective mass difference between states, and introduces off-diagonal terms into the neutrino mixing matrix, which serve to further enhance vacuum oscillation.  In effect, the neutrino eigenstates are modified in matter, with an effective mixing angle that can take the maximal value of unity under certain conditions.
In 1986 Mikheyev and Smirnov~\cite{Mikheev:1986wj} discovered these matter-enhanced neutrino oscillations by
numerically propagating solar neutrinos through the sun while taking into account
the flavor-dependent index of refraction, a phenomenon first explored by Wolfenstein~\cite{Wolfenstein:1977ue}.   Several key papers published that same year explored this phenomenon in terms of quantum mechanical level crossing, reproducing the numerical results with analytical methods~\cite{PhysRevLett.56.1305,PhysRevLett.57.1271,PhysRevLett.57.1275}. 
The conditions for this so-called ``MSW'' oscillation are realised in the solar core for higher energy solar neutrinos, causing them to be created in the matter-modified $\nu_2$ state.    Adiabatic conversion as they propagate through the Sun results in neutrinos exiting in the vacuum $\nu_2$ state, leading to a survival probability of $\sin^2\theta_{12}$.

\begin{figure}
\centering
\includegraphics[width=1.\textwidth]{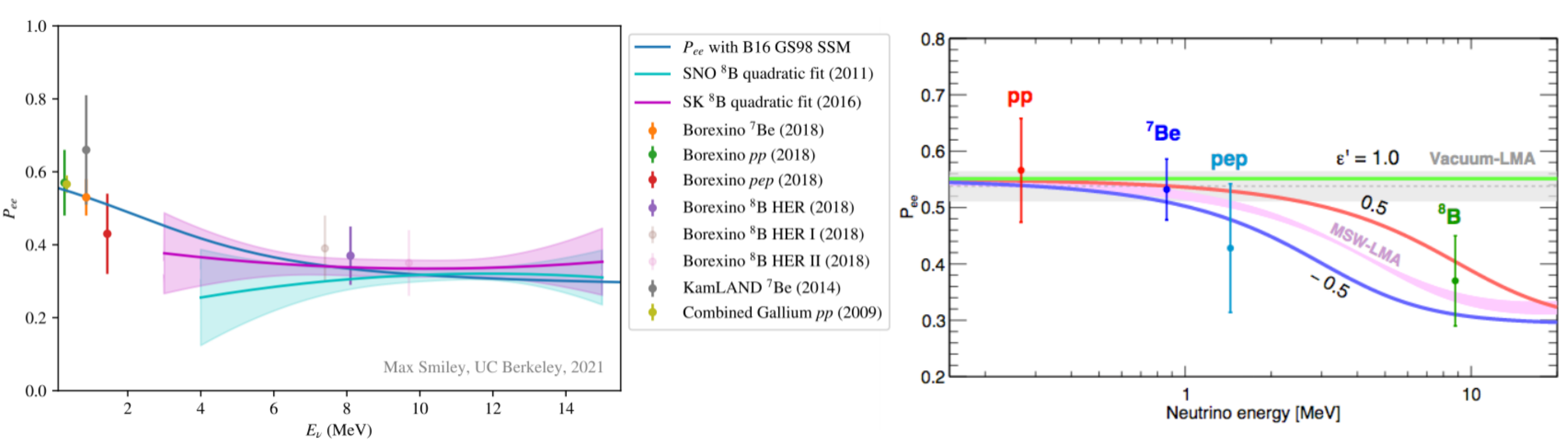}
\caption{Left: The electron neutrino survival probability curve as a function of neutrino energy (linear). Shown are the experimental data from various experiments. Figure courtesy of M. Smiley, UC Berkeley. Right: Same kind of plot (now logarithmic) showing the  survival probability as before but now for some representative curves for the NSI parameter $\epsilon'$ for values of 1.0, 0.5 and - 0.5 respectively. Also shown are the vacuum-LMA and MSW-LMA curves (from \cite{2020JHEP...02..038B}).\label{f:survivalcurve}
}
\end{figure}

The effect of MSW oscillation can be observed in the electron neutrino survival probability curve (Fig.~\ref{f:survivalcurve}).
At low energies, less than about 2 MeV, vacuum oscillations dominate, while matter effects dominate at energies above about 5 MeV.
Between these two regimes is the so-called ``transition region'', with limited data. This region is in fact the most sensitive to potential non-standard physics (Section~\ref{s:nsi}), due to the level-crossing phenomenon that arises due to the different effective masses for $\nu_e$ and $\nu_{\mu/\tau}$ in matter.
The lowest measured high energy data point suggests a certain upturn at the lower part of the curve, but this is not statistically significant. 
The uncertainties on most measurements are on the order of several percent. Hence, the constraints allow for various shapes of the survival probability curve and, thus, also for potential interesting physics. 
These possibilities open a wide field for experimental exploration.  Probing the transition region is one focus of the future experimental program, discussed in more detail in Section~\ref{s:expt}.

At higher energies (above approximately 10~MeV), a distortion is predicted in the spectrum due to regeneration of $\nu_e$ as they traverse the Earth during the nighttime.  This is termed the day/night effect, and has been sought by several experiments but, to date, remains elusive due to the small nature of the predicted effect (1--3\%~\cite{HK}).  The significance of the predicted effect depends critically on the value of $\Delta$m$^2_{12}$ -- smaller values result in a larger MSW effect in the Earth and, hence, a larger day/night asymmetry.  This offers one possible handle on a currently small discrepancy between the measured values of $\Delta$m$^2_{12}$ in solar data, and in KamLAND's terrestrial reactor data~\cite{DEGOUVEA2020135751}.  Seasonal variations in the solar neutrino flux have also been observed, and are consistent with the eccentricity of the Earth's orbit, with no evidence for additional vacuum oscillation effects~\cite{PhysRevD.72.052010}.

Almost all the solar fusion reactions, except the hep-flux, have been observed (Sect.~\ref{s:models2}). The hep flux has the highest energy but a low flux, and extends beyond the $^8$B spectrum in only a small window. It also interferes with attempts to observe the diffuse supernova neutrino background (DSNB), which is an interesting study in its own right. 
A one sided limit of the hep flux of $\Phi_{hep} < 30 \times 10^3 cm^{-2}s^{-1}$ is given by the SNO experiment \cite{2020PhRvD.102f2006A}. 

\subsection{Non-Standard Interactions}\label{s:nsi}
As most results of the neutrino measurements have uncertainties of several percent this offers opportunities beyond the Standard Model. A first topic to explore the survival curve is the introduction of Non-Standard interactions (NSI) \cite{1999PhRvD..59i3005B,2004PhLB..594..347F}. At low neutrino energies a four-fermion interaction vertex can be described by
\begin{equation}
{\cal{L}}^{NSI} = -2 \sqrt{2} G_F (\bar{\nu}_\alpha \gamma_\rho \nu_\beta) (\epsilon_{\alpha \beta}^{f\tilde{f}L} \bar{f}_L\gamma^\rho{\tilde{f}_L} + \epsilon_{\alpha \beta}^{f\tilde{f}R} \bar{f}_R \gamma^\rho\tilde{f}_R) + h.c.
\end{equation}
Here the $\epsilon$ terms denote the strength of the NSI between
the neutrinos $\nu$ of flavors $\alpha$ and $\beta$ and the left-handed 
(right-handed) components of the fermion {\it f}. In this way experimental data are needed to constrain all these parameters, some of which have already been excluded by former experiments
\cite{2013RPPh...76d4201O}. 

\begin{marginnote}[]
\entry{NSI}\\
{Non-Standard Interactions}
\end{marginnote}

Results from Borexino constrain the value of these parameters~\cite{2020JHEP...02..038B}. Future sensitivity studies have been performed for a combination of the three experiments Hyper-Kamiokande, DUNE and MICA\cite{2020PhRvD.102c5024B} as well as for the DUNE near detector \cite{2020arXiv200510272G}. 
As the $\epsilon'$ parameters in first order are mostly degenerate, studies have been performed to disentangle these parameters \cite{2020JHEP...09..106D}.

Searches for solar anti-neutrinos have been performed by the SNO ~\cite{2004PhRvD..70i3014A}, Super-Kamiokande~\cite{2020arXiv201203807S} and, more recently, Borexino experiments~\cite{BELLINI2011191}.
A  potential source is the flux from the $^{40}$K decay within the Sun, which produces an anti-neutrino flux of about 200 $\bar{\nu_e} cm^{-2}s^{-1}$ on Earth. However, the terrestrial $^{40}$K background is overwhelming. Above 3.2 MeV, the upper threshold for $^{40}$K decay, photo-fission processes in the Sun can provide higher energy anti-neutrinos \cite{1990ApJ...352..767M}. Non-standard physics processes also produce $\bar{\nu_e}$ and in turn they can be constrained by measurements \cite{2020arXiv201202142H}.   

Another source that has been debated for a long time are the highly energetic solar flares, which might produce pions in the solar atmosphere and thus produce neutrinos \cite{2004JHEP...06..045F}.

Several other non-standard physics effects can affect the shape of the $\nu_e$ survival probability in the transition region.  

Mixing of the three known neutrino states with a light sterile neutrino could modify the shape of the spectrum in the transition region, due to an additional suppression of the $\nu_e$ survival probability caused by the addition of a new mass state, $m_4$ 
~\cite{deHolanda:2010am,2020PhR...884....1D}.  Improved precision in measurements of the $^8$B spectral shape and the \emph{pep} flux would be most sensitive to this effect.  
Sterile neutrinos, like  right-handed singlets in the Standard model, would allow for significant extensions of the neutrino sector. For symmetry reasons, three singlet states could be imagined. This would considerably extend the mixing matrix. 
A neutrino magnetic moment would modify the ES cross section, with an enhancement at low energies.  
The $^7$Be line source and the spectrum of low energy neutrinos are useful probes of this effect.
A magnetic moment can be created by a one loop diagram resulting in \cite{1977PhRvD..16.1444L}:
\begin{equation}
\mu_\nu = \frac{3eG_F}{8 \sqrt{2} \pi^2}
m_\nu = 3.2 \times 10^{-19} ( \frac{m_\nu}{eV})\mu_B .
\end{equation} 
Experimental searches are based on $\bar{\nu_e}$-electron scattering searches at reactors and from astrophysical processes, especially the energy loss of RGB stars (see Section~\ref{subsec:unknown}). 
The current best limits from solar neutrinos come from Borexino's studies of $^7$Be~\cite{arpesella:2008}, and the low-energy solar neutrino spectrum~\cite{2017PhRvD..96i1103A}.

The SNO experiment has produced limits on the lifetime of the the second neutrino state ($\nu_2$) from a study of the $^8$B spectral shape~\cite{PhysRevD.65.113009}.
Limits on neutrino lifetime also exist from cosmological studies~\cite{2020arXiv201101502B}.

\subsection{Emission of (un-)known particles \label{subsec:unknown}}

Another area of solar physics is the search for solar axions.
or more general axion-like particles (ALPs) \cite{2015ARNPS..65..485G,RevModPhys.93.015004}. 
The axion is a pseudoscalar object like the $\pi^0$. The axion can decay into 2 photons via a triangle graph with a coupling constant $g_{a\gamma}$. In a crossed Feynman diagram where one photon is provided from an external electromagnetic field, a mono-energetic $\gamma$ -line will be generated (Primakoff-effect). The axion can be searched for in stellar objects like the Sun via the Primakoff effect $\gamma + Z e^- \rightarrow a + Ze^-$. Axions may play an important role in cosmology. 
Based on the Primakoff process, constraints on $g_{a\gamma}$ from astronomical objects have been deduced and now experimental observations are split in two groups: haloscopes to search for cold dark matter in the Milky Way and helioscopes searching for a thermal solar axion flux.

A first limit on axion parameters can be deduced from astrophysics using the age of the Earth. Currently, the Sun is about halfway through its main sequence evolution. Hence, the solar axion luminosity must not exceed its photon luminosity, otherwise its nuclear fuel would have been spent before reaching the current age of the Sun. This requirement puts constraints on the coupling $g_{a\gamma} \leq 2.4 \times 10^{-9}GeV^{-1}$. The solar axion spectrum has been calculated in \cite{1987PhRvD..36.2211R,1990PhR...198....1R}.
This provides an energy region of interest for experimental searches between 1-9 keV, i.e the X-ray region.

As with solar neutrinos, experiments searching for axions are often performed underground. A number of such searches have been performed~\cite{2015ARNPS..65..485G}, 
 and further experiments are in preparation or under consideration to explore the parameter space \cite{2016PhR...643....1M}.
 Independent constraints come from astrophysics,for example from helioseismology \cite{1999APh....10..353S,vinyoles:2015}. Most of these studies are based on energy loss arguments \cite{1990PhR...198....1R,2020PhRvD.102h3007C}, in which axions add another component of energy loss in stars. This is especially important at the red giant branch (RGB) \cite{2019arXiv191010568A}. Additional stellar energy losses will affect the Hertzsprung Russell diagram, which governs the evolution of stars. Energy loss arguments can be applied to studies of globular clusters with a decent number of stars to deduce limits on the axion~\cite{1987PhRvD..36.2211R} and can also produce
 limits for a magnetic moment of the neutrino ~\cite{2019arXiv191010568A}.

%% file: experiments.tex
\section{DETECTION OF SOLAR NEUTRINOS}\label{s:expt}

The main challenges for the next generation of solar neutrino measurements lie in a trifecta of requirements: scale, cleanliness, and depth.  The sheer size needed for solar neutrino detection has been a requirement of long standing, due to the weak nature of neutrino interactions.  The exquisite precision now demanded for further discovery places even more stringent requirements on both depth, to restrict cosmogenic muon-induced backgrounds, and cleanliness.  
It is worth noting the importance of the shape of a site's overburden in evaluating the total muon rate: a flat overburden offers significant advantage in reducing the total rate of muons. 
\begin{marginnote}[]
\entry{Solar neutrino experiments}{ require large, ultra-clean detectors located deep underground to shield from cosmic radiation.}\end{marginnote}

Neutrino detection is extremely challenging, due to their very low reaction cross sections.
ES of neutrinos on electrons is sensitive to all 3 flavours, but with a significant (approximately factor of 6.5) enhancement for $\nu_e$.  A great advantage of the ES reaction is the strong correlation between the direction of the outgoing electron and that of the incoming neutrino, giving a pointing capability.  The cross section for $\nu_e$ undergoing ES is on the scale of 10$^{-45}$ to 10$^{-43}$~cm$^2$ across the full range of solar neutrino energies, hitting 4.3$\times$10$^{-44}$~cm$^2$ at 5~MeV, with a rising energy dependence.  
The CC reaction occurs only for $\nu_e$ at energies relevant for solar neutrinos.  This reaction has a more peaked differential cross section than the very broad ES dependence, offering a more precise measurement of incident neutrino energy.  The reaction also has a weak angular correlation.  The cross section for interaction on a deuteron is approximately an order of magnitude higher than that for ES at 5~MeV, at 3.5$\times$10$^{-43}$~cm$^2$~\cite{Kubodera:1993rk}, although the number of available targets in a detector such as SNO is significantly lower than for ES.  The CC reaction on $^7$Li is almost an order of magnitude higher than that on the deuteron at 5~MeV, at 1.5$\times$10$^{-42}$~cm$^2$~\cite{Bahcall:1978fa}, although both fall off rapidly at lower energies.  A target-weighted cross section for a water, heavy water, and 10\% Li-loaded detector is presented in~\cite{TheiaWP}.  CC reactions on $^{71}$Ga~\cite{PhysRevC.56.3391} and $^{37}$Cl~\cite{Bahcall:1996qv} have been used to great effect in radiochemical experiments, although the cross sections are approximately 2 and 20 times lower than for $^7$Li, respectively, in the range 2--5~MeV.  \cite{asdc} presents a comparison of the CC reactions on Li, Cl and Ga.
A NC measurement offers flavour-blind neutrino detection.  The cross section on the deuteron is 9.5$\times$10$^{-44}$~cm$^2$ at 5~MeV~\cite{Kubodera:1993rk}, more than a factor of two higher than the ES.

Next-generation experiments will focus on the ability to make a precision determination of the CNO neutrino fluxes, referring to neutrinos produced in both the CN- and NO-cycles (Sec.~\ref{subsec:Nuclear:Intro}), which would resolve questions in solar metallicity (Sec.~\ref{s:models1}). 
Improved accuracy in the measurement of the shape of the $^8$B solar neutrino spectrum, particularly in the sensitive 1--5~MeV transition region between the low-energy, vacuum dominated regime and the higher-energy, matter-dominated regime, would allow for tests of a number of non-standard models, including flavour-changing NC interactions, and certain models for sterile neutrinos (Sec.~\ref{s:physics}).  Precision measurements of the \emph{pp}
and \emph{pep} fluxes can probe and monitor the solar luminosity (Sec.~\ref{s:models2}); $^7$Be and $^8$B can constrain the temperature of the solar core; and a measurement of the day/night asymmetry can constrain oscillation parameters, and confirm our understanding of the interaction of neutrinos with matter.  
\begin{marginnote}[]
\entry{Day/night asymmetry}{An asymmetry in the $\nu_e$ flux measured during the day vs the night, due to regeneration of $\nu_e$ during passage through the earth}
\end{marginnote}

\subsection{Detection techniques}

Due to their weak interactions, large detectors are needed to gather enough statistics for precision measurements of solar neutrinos.  Sec.~\ref{s:intro} describes the tremendously successful program of experiments that first observed solar neutrinos, demonstrated neutrino flavour change and non-zero mass and, more recently, have moved into precision spectroscopy.  These experiments fall into two main categories: radiochemical experiments, and large, monolithic optical detectors.  The former integrate over a period of many days or weeks to collect data, focusing their use on integral flux measurements.    
For real-time detection and precision spectroscopy, large optical detectors have proved to be the work horses of the field.

\subsubsection{Radiochemical detectors}

As discussed in Section~\ref{s:intro}, radiochemical experiments were critical in the understanding of the solar neutrino problem.  The SAGE experiment is still operating, although largely focused on the so-called ``gallium anomaly'', relating to a deficit in $\nu_e$ observed from electron capture sources ~\cite{2019sone.conf...29G,Kostensalo:2019vmv}.  Radiochemical detection is still an area of active and ongoing research.

Perhaps the isotope with the lowest threshold for solar neutrino detection is $^{205}$Tl, with a Q-value of only 52 keV, which triggered the idea for the LOREX experiment several decades ago \cite{2018NIMPA.895...62P}. The daughter $^{205}$Pb is very long-lived, with a half life of several million years and, thus, would not form a coincidence, but it would be an almost unique opportunity to measure the average solar neutrino flux over the last million years. The LOREX experiment is actively working on this radio-chemical approach. Recent new nuclear cross section calculations on this reaction are predicting lower solar neutrino interaction rates than in the past \cite{2020PhRvC.101c1302K}. The bound state beta decay was recently observed for the first time at the GSI research centre in Germany, which is an essential ingredient for such a potential $^{205}$Tl measurement. 
Another nuclide that could provide information of the averaged neutrino flux over the last million years is $^{98}$Mo \cite{1985AIPC..126..196W}.

\subsubsection{Real-time detectors}

The large, monolithic optical detectors used for real-time observations use either a water target or a scintillating liquid.  In both cases, the target produces light in response to the passage of charged particles.
Pure liquid scintillator (LS) detectors offer high light yields, resulting in the extremely good energy resolution and low thresholds critical for addressing the vacuum-dominated regime of solar neutrino oscillation.  They can also achieve impressively low levels of radioactive contamination, with the Borexino experiment repeatedly paving the way with new standards in cleanliness.  These ultra-clean, high light yield detectors can target the lowest energy solar neutrino fluxes and spectra, including the CNO, pep, and even pp neutrino fluxes.  Water Cherenkov detectors offer the benefit of directional resolution for background rejection.  The sheer volume of detector that can be constructed, thanks to the excellent attenuation lengths achievable with ultra-pure water, can provide unprecedented statistics, offering insights into both the $^8$B spectral shape, and day/night asymmetry.  These two detector types are highly complementary, addressing opposite ends of the spectrum of solar neutrino physics.

In more recent years, a number of new developments could facilitate a new kind of experiment, with the capability to answer many of the critical open questions discussed in Sections~\ref{s:models1} and~\ref{s:physics}.  
Perhaps the most promising avenue for next-generation detectors is the concept of a hybrid optical neutrino detector, capable of leveraging both Cherenkov and scintillation signals in a single detector.  This is an extremely challenging, but potentially ground-breaking development, which could enable a new generation of detectors with world-leading sensitivity across a broad range of physics goals.  For solar neutrinos, this would enable directional detection to low thresholds, as well as additional particle identification capabilities from the Cherenkov/scintillation ratio and the time profile of detected light.  This could significantly improve background rejection and signal efficiency.

This hybrid detection can be achieved in a number of ways: 
\begin{itemize}
    \item by deploying a target material that modifies the scintillation signal in a number of ways, either by reducing the intensity~\cite{BNLwbls1,BNLWbLS3,WbLSother,Onken} or by delaying the time profile~\cite{MinfangSlow,BillerSlow}, in order to enhance separation from the fast, lower-intensity Cherenkov signal;
    \item by deploying fast photon detectors~\cite{LAPPD} to help differentiate prompt Cherenkov from the typically slower scintillation light;
    \item or by utilizing spectral sorting to separate the Cherenkov and scintillation signals by their wavelength~\cite{dichroic2}.
\end{itemize}
Substantial work has been dedicated to realising this concept, both experimentally~\cite{ChineseChS,CHESSwbls,CHESS2,CHESS1,Flatdot}, and in development of new analysis tools to leverage and enhance this simultaneous detection for large detectors~\cite{GGimaging,BillerMulti,AberleDirection,WonsakTracks,ElaginST,AoboCNN,LSresponse}.

\begin{figure}
    \centering
    \includegraphics[width=0.85\textwidth]{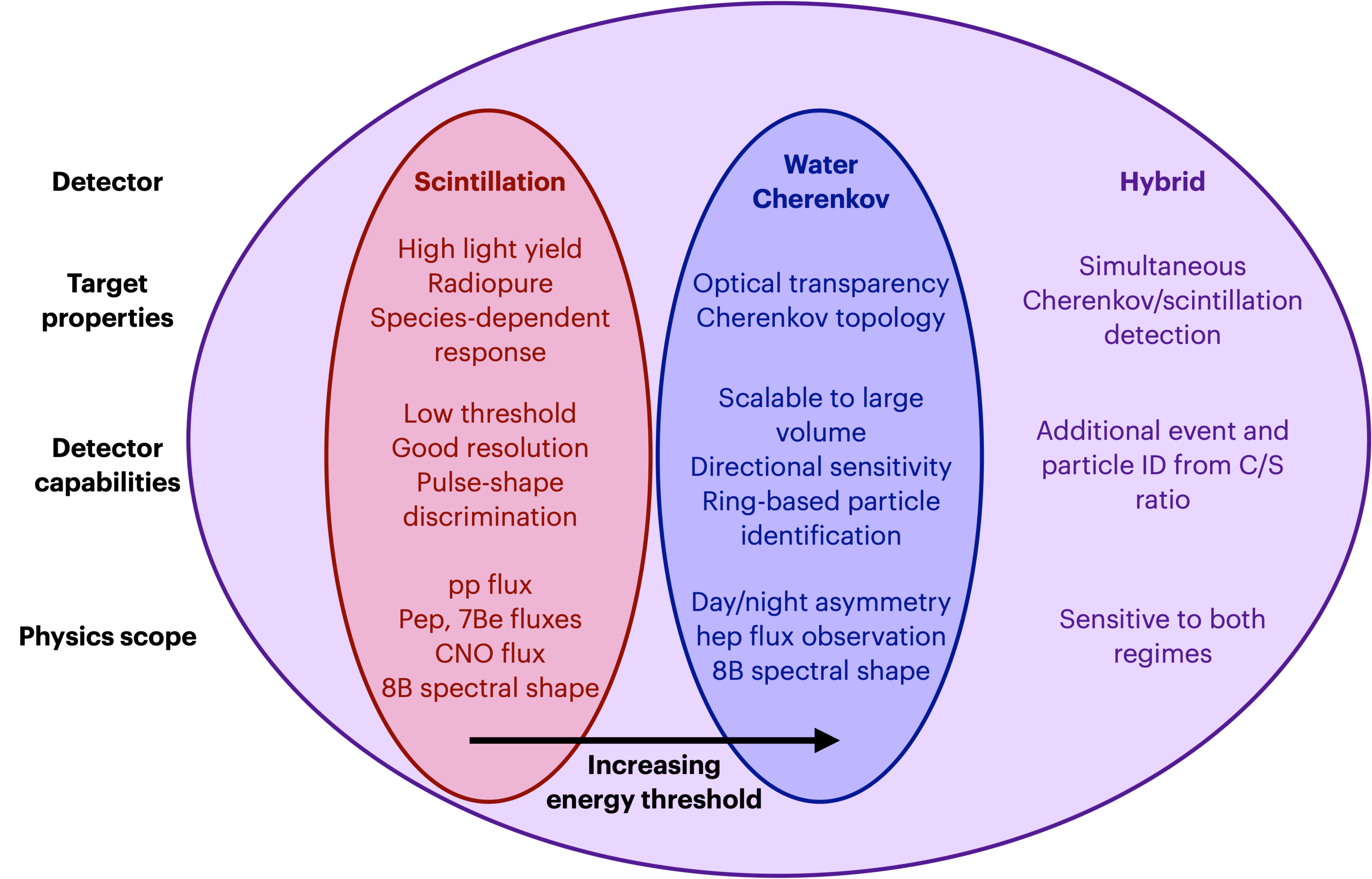}
    \caption{Conceptual illustration of the benefits offered by different real-time solar neutrino detection techniques, and the physics they can address.}
    \label{f:venn}
\end{figure}

The optimal configuration will depend on the exact detector geometry.  For example, the long-wavelength tail of Cherenkov light travels more quickly than the predominantly blue scintillation.  Thus, in a large detector dispersion effects can serve to enhance time-based signal separation; in a smaller detector, in which dispersion effects play less of a role, deployment of fast photon detectors may be more critical.  The optimal configuration for any particular detector would involve optimization along all the above axes, and will likely also be at least in part determined by local factors such as the practicalities of underground deployment of scintillator, readout requirements for different photon detector choices, and the requirements of other physics goals, which may place a premium on high light yields or, conversely, on a high-fidelity Cherenkov signal.
Figure~\ref{f:venn} shows the benefits offered by each of these detector types, and the solar neutrino physics they can address.

\subsubsection{Isotopic loading}

Isotopic loading can  offer enhancements for solar neutrino detection, for example, by offering a CC detection channel, such as that used in SNO, which provides greater precision on the underlying neutrino spectrum.  This may be important for probing details of the shape of the $^8$B spectrum in the transition region between matter- and vacuum-dominated oscillation, for example.  A number of techniques are being explored to load large target masses while retaining the good optical properties critical for precision low-energy physics~\cite{nano,MinfangLoading}.  Candidates include $^7$Li, which has a favourable cross section in the energy range relevant for probing the transition region.

$^{100}$Mo is a candidate target for real-time measurements of  pp neutrinos, with a low threshold and favourable cross section, and the signal has a nice coincidence with less than one minute. A similar time coincidence can be formed from the DBD isotope $^{116}$Cd \cite{2003PhLB..571..148Z}. It is used in various NLDBD experiments, but solar pp detection is not possible as the Q-value is about 460 keV, just above the pp-neutrino flux, making these two spectra difficult to disentangle. 
DBD isotopes with Q-values of more than 1 MeV exist, and provide a potential target for solar pp-neutrinos, especially $^{150}$Nd \cite{2012PhLB..709....6Z}.

A further potential nucleus for a pp-measurement is $^{115}$In. This nuclide was investigated in the LENS project~\cite{2000NuPhS..87..195C}, but activity waned due to the challenge of achieving the required low background.

\subsection{Water Cherenkov experiments}
\subparagraph{Sudbury Neutrino Observatory and Super-Kamiokande}
Water Cherenkov experiments were critical in the resolution of the solar neutrino problem.  The first indication of a solution came from a combination of the CC measurement from SNO~\cite{snocc}, sensitive only to the $\nu_e$ flux, with the high statistics ES measurement from Super-Kamiokande (Super-K)~\cite{PhysRevLett.86.5651}, which is sensitive to all 3 flavours, with an enhancement for the electron flavour.  These two measurements disagreed at the 3$\sigma$ level, demonstrating the presence of some non-electron flavour neutrinos in the flux from the Sun.  SNO's seminal NC measurement, sensitive equally to all 3 flavours, confirmed that the solar neutrino problem was due to the $\nu_e$ produced in the solar core changing flavour prior to detection on the Earth~\cite{2004PhRvL..92r1301A}. 

Both experiments have since improved the precision of $^8$B flux and oscillation parameter measurements, as well as pushing to low energy thresholds in order to probe the shape of the $^8$B spectrum in the MSW transition region~\cite{leta,2013PhRvC..88b5501A,2016PhRvD..94e2010A}.  To-date, the measured spectral shape is consistent with the MSW-predicted upturn, but also with several non-standard effects (Sec.~\ref{s:physics}). Both experiments have sought evidence of the day/night effect~\cite{PhysRevLett.112.091805,snodn}.  Although Super-K initially saw nearly 3$\sigma$ indication of this effect, subsequent data were seen to reduce the significance. 
Greater statistics are required to confirm this effect.  
Future reactor data at Super-K may allow sufficient precision on $\Delta$m$^2_{12}$ to resolve the current small discrepancies between reactor and solar data~\cite{DEGOUVEA2020135751}.
A search for \emph{hep} neutrinos is extremely challenging due to the very low predicted flux.  The best limits on this flux come from SNO, and are currently a few times the SSM prediction~\cite{2020PhRvD.102f2006A}.

SNO ceased operation in 2006.  A number of more recent analyses of the data set have provided constraints on non-standard effects such as Lorentz violation, and neutrino decay~\cite{SNOLV,SNOdecay}.  
Super-K continues to take data, and will continue to improve on both statistics and precision for the $^8$B measurements.  Super-K has recently been upgraded with the addition of gadolinium, a project known as SK-Gd, which will enhance neutron capture efficiency~\cite{PhysRevLett.93.171101,Takeuchi:2020slv}.

\subparagraph{Hyper-Kamiokande}
The Hyper-Kamiokande (Hyper-K) experiment is expected to start construction and operation this decade~\cite{HK}.  At over 250 ktonne total mass, with 40 \% coverage, 
Hyper-K has the potential to contribute a great detail to the picture of high-energy solar neutrinos, in particular.  

The small day/night effect and the relatively flat spectrum measured by SNO and Super-K serve to define the values of the oscillation parameters, and result in some small tension with terrestrial measurements at the KamLAND experiment.  Within 10 years of operation, Hyper-K will be able to measure the day/night asymmetry to better than 4 (8) $\sigma$ at the values currently predicted by reactor (solar) experiments.  Hyper-K also expects improved sensitivity to both the $^8$B shape and \emph{hep} neutrinos due to sheer size, and improved light collection efficiency relative to Super-K.

\begin{marginnote}[]
\entry{Water Cherenkov experiments} {These easily scalable detectors offer the potential to observe the day/night asymmetry, hep neutrinos, and to probe the shape of the $^8$B spectrum.}
\end{marginnote}

\subsection{Liquid scintillator experiments}

\subparagraph{Borexino}
The high light yield of organic scintillators offers high precision spectroscopy as well as low thresholds.
The Borexino collaboration has laid new ground for scintillator-based detection of solar neutrinos, with world-first direct measurements of pp~\cite{2014Natur.512..383B} and pep neutrinos~\cite{Borexino:pep}, the best precision of $^7$Be~\cite{arpesella:2008,Borexino:7Be}, as well as $^8$B flux and spectrum measurements.  This comprehensive spectroscopic study of the pp-chain of solar neutrinos~\cite{Borexino:phaseI,borexino:pp2,borexino:pp3} is complemented by the first detection of neutrinos from the sub-dominant CNO cycle~\cite{2020Natur.587..577B}, a ground-breaking achievement.  This experiment will cease taking data shortly, but it has paved the way for a new generation of scintillator experiments to follow.  One of Borexino's particular accomplishments was the astonishing level of radioactivity purity achieved in the detector: levels approaching $10^{-19}$g/g of both $^{232}$Th and ${238}$U and, most critically for the CNO measurement, a rate of $^{210}$Bi events $\leq 11.5\pm 1.3$~counts per day per 100 tonnes of detector material (cpd/100t).  This can be compared to the fitted CNO rate of $7.2^{+3}_{-1.7}$ cpd/100t.  The main limiting factors to improved precision remain lingering sources of radioactivity, in particular even these low levels of $^{210}$Bi, and  cosmogenic-induced backgrounds such as $^{11}$C.  A leading consideration for improved precision in the \emph{pp} measurement is the $^{14}$C background, inherent in any organic scintillator.

\subparagraph{KamLAND}
The KamLAND detector has a long history of discovery, including the seminal paper that demonstrated that the neutrino flavour change observed by SNO was in fact due to oscillation~\cite{2003PhRvL..90b1802E}.  Reactor measurements of the $\theta_{12}/\Delta m^2_{12}$ 
sector of oscillations  with KamLAND data provide a terrestrial comparison to solar neutrino results, with some small tension persisting in the value of $\Delta m^2_{21}$.  The KamLAND Collaboration have also made several measurements of solar neutrinos, including a measurement of the spectral shape of $^8$B neutrinos at low energy~\cite{PhysRevC.84.035804}, and detection of $^7$Be neutrinos~\cite{PhysRevC.92.055808}.  Since then, the KamLAND detector has been upgraded for a world-leading NLDBD search~\cite{KamLAND-Zen:2016pfg,KamLAND-Zen:2019imh}, with an inner containment vessel deployed in the centre of the detector, containing Xe-loaded scintillator.  This detector continues to be sensitive to solar neutrinos.  Future measurement potential includes $^8$B flux and spectral measurements, as well as more exotic analyses such as searches for solar antineutrinos.

\subparagraph{SNO+}
SNOLAB, in Ontario, Canada, offers one of the deepest sites available worldwide for low-background studies, at a depth of 6 km water equivalent.  This results in incredibly low cosmogenic backgrounds, in particular the $^{11}$C that can be a limiting factor in precision  low-energy solar neutrino measurements.
The SNO detector has been converted from a water Cherenkov detector to a pure LS detector, as part of the programme for the SNO+ experiment~\cite{snopLS}.  The primary goal for SNO+ is a search for NLDBD via loading of the LS with tellurium. Like KamLAND-Zen, this detector will have sensitivity to solar neutrinos, in particular the $^8$B neutrinos that lie above the 2$\nu\beta\beta$ decay endpoint of $^{130}$Te (approximately 2.5 MeV)~\cite{Andringa:2015tza}.   SNO+ will be the deepest low-background neutrino experiment operating and, with a mass of 780 tons of scintillating target, a future phase of the experiment has the potential to contribute across the breadth of solar neutrino physics.  

Early data from the initial water phase of SNO+ have already demonstrated low levels of cosmogenic and external background, allowing a measurement of the $^8$B spectrum in water~\cite{2019PhRvD..99a2012A}.  LS data should allow improved precision, and a significantly lower threshold.  
Preliminary LS data from SNO+ shows levels of radon daughters 
in the detector that would make low-energy solar neutrino measurements challenging. 
Precision measurements of this regime would require significant reduction of these backgrounds, similar to that achieved during the first period of Borexino operations~\cite{arpesella:2008}. This could potentially be addressed via an extensive campaign that would include recirculating through the SNO+ scintillator process systems, built with several purification capabilities, along with other background reduction techniques.

Estimates of the SNO+ sensitivity to CNO neutrinos can be made under the assumption of certain levels of background reduction.  The radiopurity levels observed in early LS data already meet the targets for the primary goal of NLDBD, and might be sufficient to permit a limited-precision measurement of the CNO flux, with large uncertainties due to the presence of backgrounds.  With significant further background reduction, of approximately a factor of 10 for the U- and Th-chains and 1000 for $^{210}$Bi, negligible $^{40}$K, 
and constraining the \emph{pep} flux based on the \emph{pp} flux, as was done by Borexino in their discovery paper, SNO+ could achieve better than 15\% precision on the CNO flux.

SNO+ will also offer improved precision in $\Delta m^2_{12}$.  Sensitivity to this parameter using both solar neutrinos and reactor neutrinos will provide additional data to resolve current (small) discrepancies in measurements from solar and terrestrial sources.  

\subparagraph{JUNO}
The Jiangmen Underground Neutrino Observatory (JUNO) is due to start construction in 2021, with data taking to follow within a few years~\cite{JUNO,JUNOCDR}. At 20 ktonne, with approximately 75\% coverage, a goal of 3\% energy resolution at 1~MeV, and a target of $10^{-17}$g/g intrinsic $^{238}$U and $^{232}$Th, JUNO will be a ground-breaking achievement in low-energy neutrino detection.  JUNO's primary goal is measurement of the neutrino mass hierarchy using reactor neutrinos.  The relatively shallow overburden of 680~m limits the low-energy solar neutrino program.  However, at the target background levels, a threshold of 2~MeV could be achieved for $^8$B neutrinos.  This is substantially lower than that achievable in a Cherenkov detector, and 1 MeV lower even than achieved by Borexino for a measurement of the $^8$B spectral shape~\cite{PhysRevD.82.033006}.  Combined with the large volume, which results in rapid collection of statistics, JUNO will have sensitivity to non-standard interactions that could affect the $^8$B spectral shape, and 2 (3) $\sigma$ sensitivity to the day/night effect at current reactor- (solar-) favoured parameter values~\cite{JUNO8B}.  With the ability to measure $\Delta m^2_{12}$ to percent-level precision from reactor neutrinos, and to approximately 20\% using solar neutrinos, JUNO will provide a uniquely precise cross check on the consistency of data from these two sources.  Precision measurements of the $^7$Be and $^8$B fluxes are also possible. 

\begin{marginnote}[]
\entry{Organic scintillator experiments} {These low-threshold detectors can probe low-energy solar neutrinos, including pp and pep neutrinos, the $^8$B transition region and, if sufficiently clean, the CNO neutrino flux.}
\end{marginnote}

\subsection{Hybrid optical neutrino detectors}

\subparagraph{Jinping}
The Jinping underground laboratory in China~\cite{JinpingLab} has a 2.4-km rock overburden, resulting in a cosmic-ray muon flux almost as low as that at SNOLAB~\cite{JinpingMuons}. 
A 5-ktonne scintillator detector is planned at this site, with the goal of deploying a high-light yield organic scintillator, with careful target selection and high-precision, high-coverage instrumentation  such that the Cherenkov signal can be leveraged for direction reconstruction and particle identification. 
Data from a 100-tonne-scale prototype could be available as early as 2024.  The depth, size, and hybrid optical detection could allow for unprecedented precision in solar neutrino measurements.  The Jinping  Neutrino Experiment would have sensitivity across the full spectrum of solar neutrinos.  The Letter of Intent explores the capabilities for a range of detector sizes and light collection value, from 1- to 4-ktonne in fiducial mass, and 200 to 1000 photoelectron/MeV light collection~\cite{JinpingLOI}.  The highest performing detector under consideration would achieve percent-level measurements of the \emph{pp}, $^7$Be, and $^8$B neutrinos, as well as a few-percent uncertainty on the \emph{pep} neutrino flux.  A CNO measurement is highly dependent on both target mass and resolution, but could reach better than 15\% precision for the larger, high-resolution detector configurations.  The experiment would also have good sensitivity to probe the $^8$B spectral shape. 

\subparagraph{\textsc{Theia}}
 \textsc{Theia}  is a proposed large-scale hybrid optical neutrino detector~\cite{TheiaWP}, which is a realisation of the Advanced Scintillation Detector Concept first proposed in~\cite{asdc}.  Considering both WbLS and other novel LS materials and isotopic loading options, along with cutting-edge photon detection technology, \textsc{Theia}'s proposed site at the Sanford Underground Research Facility (SURF)  laboratory in SD, USA would offer a high-energy neutrino programme as part of the Long Baseline Neutrino Facility, as well as a broad program of low-energy physics.  Although pure LS offers improved radiopurity, directional sensitivity would at least in part offset the increased levels of contamination inherent in the water component of a WbLS target.
 At its full, 50 to 100-ktonne size, \textsc{Theia} could achieve better than 10\% precision on CNO solar neutrinos with a WbLS target, or percent-level precision with pure LS~\cite{Bonv,TheiaWP,MeVSolar}.  This would allow a high-confidence resolution of the metallicity problem.  Possible isotope loading is also being explored to offer a CC interaction, which would provide a high fidelity measure of the underlying neutrino spectrum, potentially offering enhanced sensitivity both to CNO neutrinos, and to the $^8$B spectral shape.

\begin{marginnote}[]
\entry{Fluor}{A material added to scintillator that shifts the spectrum of emitted light to a region chosen to enhance photon propagation. 
Fluors can affect both the number and time profile of emitted photons.}
\end{marginnote}
Even a small, few-hundred ton hybrid detector could offer enticing reach for solar neutrinos if filled with a high light yield material, as explored in~\cite{MeVSolar}.  This paper assumed a standard scintillator mixture of linear alkyl benzene (LAB) loaded with 2g/L of the fluor, PPO, and studied the impact of different photon detectors.  With current standard photomultiplier tubes, with 1.6-ns transit time spread, such a detector could achieve 14\% precision on the CNO neutrino flux, dropping below 10\% if a constraint is imposed on the pep flux based on knowledge of the pp flux, as done by Borexino in their discovery paper~\cite{2020Natur.587..577B}.  With fast photon detectors, such as LAPPDs~\cite{LAPPD}, the precision is below 5\%.  
Slow fluors, 
such as those studied in ~\cite{MinfangSlow,BillerSlow}, can further enhance the separation of the prompt Cherenkov signal by delaying the scintillation light, and may offer further performance improvements, although this effect must be balanced against potential degradation in vertex reconstruction due to reduced precision in photon time-of-flight information.
\begin{marginnote}[]
\entry{PPO}{2,5-diphenyhloxazole.  2g/L of PPO enhances the light yield of LAB by approximately an order of magnitude, and shortens the time profile significantly, resulting in improved vertex reconstruction.}
\end{marginnote}

\subparagraph{Other hybrid detectors}
A group in Korea is exploring the option for a few-ktonne scale detector in the Yemilab~\cite{Korea}.  
Both water-based liquid scintillator and pure scintillator are under consideration.

\begin{marginnote}[]
\entry{Hybrid Cherenkov/scintillation detection} {A hybrid optical detector, capable of both Cherenkov and scintillation detection, would allow improved background rejection via directional reconstruction, and particle identification based on the Cherenkov/scintillation ratio.  This could improve precision for both low- and high-energy solar neutrino physics.}
\end{marginnote}

\subsection{Noble liquid and solid-state experiments}

As experiments continue to break new ground, solar neutrinos can become a background for other searches, which offers new opportunities for discovery. 
Experiments designed to search for coherent neutrino-nucleus elastic scattering, known as CE$\nu$NS, will be sensitive to higher-energy solar neutrinos via this channel. Proposed many decades ago~\cite{cevns0,cevns02,cevns03}, first detection of CE$\nu$NS interactions was achieved recently at Oak Ridge~\cite{cevns,cevns2}. The cross section for CE$\nu$NS interactions is favourable due to an A$^2$ dependence; however, the nuclear scatters typically fall below 10~keV, requiring detectors with excellent resolution and very low threshold. 
Next-generation noble liquid dark matter experiments, designed to search for nuclear recoils from WIMP interactions, will be sensitive to the so-called ``neutrino floor'', where solar neutrinos, amongst other sources, have the potential to become a dominant background~\cite{DMreview}. Sensitivity in these detectors comes via two channels: CE$\nu$NS  above 5~MeV, and elastic scattering in the keV to MeV range.  

\subparagraph{Solid-state detectors}

A combined CE$\nu$NS and ES signal in a ton-scale Ge detector would offer improved sensitivity to the shape of the survival probability curve across the range from vacuum- to matter-dominated oscillation, and give sensitivity to possible active-to-sterile mixing~\cite{sterile1409}.

Extremely low-threshold detectors such as SuperCDMS could potentially detect \emph{pp} and other low-energy branches via this route, if thresholds in the few-eV range can be achieved~\cite{supercdms,nucleus}.  Equaling the precision of the Borexino measurements would require 500 (5000) kg.yr exposure for \emph{pp} and $^7$Be (pep and CNO) neutrino fluxes~\cite{LouisLowth}.

\subparagraph{Noble liquid detectors}

An advantage of these inorganic scintillating materials is the lack of intrinsic $^{14}$C background, which forms a dominant background to \emph{pp}-flux measurements in organic scintillators.  Noble liquid detectors also offer excellent discrimination between electron- and nuclear-recoil signatures.

A large LXe detector would have sensitivity to $^8$B neutrinos in the 5--15~MeV range via CE$\nu$NS, which could offer a sensitive test of the spectral shape in this region, probing the possible presence of NSIs, as well as allowing discrimination between the presence of NSIs and a possible dark-side solution for oscillation parameters ($\theta_{12}>45^{\circ}$~\cite{LNG:NSI,cevns:nsi,cevns:darkside,cevns:darkside2,cevns:nsi2,cevns:nsi3}.  
A search for such events at XENON1T did not observe a signal, but shows promise for a discovery in the forthcoming XENONnT experiment~\cite{Aprile:2020thb}.
Improved resolution in the few-keV nuclear recoil energy range could permit discrimination and first discovery of the \emph{hep} neutrinos. 

Electron recoils from ES interactions give rise to signatures of tens to hundreds of keV.   High light yield and excellent energy resolution would allow studies of \emph{pp} and other low-energy, high-flux branches with good precision via this channel. 
A percent-level ES measurement of \emph{pp} neutrinos may be possible in a large LXe detector such as DARWIN~\cite{solar:dm,DARWIN,2020EPJC...80.1133D}, which aims for 50 tons of LXe.  This would require either depletion of $^{136}$Xe, or to focus on the very low-energy regime where the 2$\nu\beta\beta$ spectrum falls rapidly. This may also offer a path to the lowest energy measurement of $\sin^2\theta_W$, if one independently imposes the luminosity constraint.  This assumes significant reductions in radiopurity beyond current-generation detectors, of nearly 3 orders of magnitude, to approximately 10 events /ton/yr/keV in the electron recoil band.  Another background would come from neutrino capture on $^{131}$Xe (21\% abundance), which would add events in the region from 355-420 keV, relevant for a pp measurement. New nuclear cross section calculations show increased solar neutrino interaction rates~\cite{2020arXiv200901164K}.
If the $2\nu\beta\beta$ spectrum could be reduced by 3 orders of magnitude, a measurement of CNO neutrinos could be possible~\cite{LouisCNO}.

The projected 1,000 tonne-year exposure of a next-generation two-phase liquid argon (LAr) TPC would collect over 10,000 CNO ES events, and offer sensitivity to $^7$Be and \emph{pep} neutrinos.  Flux measurements would require radon reduction on the order of 3 orders of magnitude beyond that measured in smaller detectors, to $200 \mu$Bq of $^{222}$Rn/100t, equivalent to 16~cpd/100t, although this is expected to be at least partially mitigated by the additional shielding of a larger detector.  This can be compared to a predicted CNO rate above threshold of 0.64 (0.90) cpd/100t for low- (high-) Z models. 
Underground argon (UG-Ar) is also required, 
to limit the impact of other backgrounds.  DarkSide-50 demonstrated levels of $^{39}$Ar in UG-Ar over $10^3$ lower than atmospheric argon, at 0.7~mBq/kg~\cite{PhysRevD.93.081101}.  
$\beta$ decay of $^{42}$K, a daughter of $^{42}$Ar, has an end-point 3.52~Mev and contributes a potentially significant background in the region of interest.  Measurements of atmospheric argon show $^{42}$Ar activity of 94.5$\pm$18.1~$\mu$Bq/Kg~\cite{Cattadori_2012}, equivalent to $\sim 8\times10^5$cpd/100t.   
Studies assume this background would be significantly reduced by use of UG-Ar, such that it can be neglected.
With these assumptions, such a detector could achieve approximately 17 (23)\% precision on CNO for high- (low- )Z models, and a few-percent precision for $^7$Be~\cite{LArTPC}.

Detection via CC interactions has been studied for both LXe and LAr detectors. CC capture on $^{136}$Xe offers a delayed coincidence signal, which could provide a low-background approach for CNO neutrino detection, and a precision measurement of the $^7$Be energy, yielding insight on the core temperature of the Sun~\cite{LXeCC}.

The long-baseline neutrino experiment, DUNE, will have sensitivity to high-energy solar neutrinos via both ES and CC interactions in the LAr target~\cite{DUNEcdr}.  
Above 5.9~MeV, CC interactions on $^{40}$Ar result in transitions to excited states in $^{40}$K~\cite{2018PhRvC..97c4309K}, and offer good resolution for reconstructing the neutrino energy. 
While the threshold in this detector is unlikely to permit measurements of CNO neutrinos, or the $^8$B spectral shape in the MSW transition region, it may offer sensitivity to the $^8$B flux, first detection of \emph{hep} neutrinos, and sensitivity to oscillation parameters via the day/night asymmetry~\cite{BeacomDUNE}.

\begin{marginnote}[]
\entry{Noble liquid experiments} {Built primarily for rare event searches, as they move towards larger scales these detectors also offer sensitivity to solar neutrinos.  Low-energy fluxes can be detected primarily via ES, and $^8$B neutrinos via CE$\nu$NS.}
\end{marginnote}

\subsection{Prospects}
\begin{figure}
    \centering
    \includegraphics[width=0.95\textwidth]{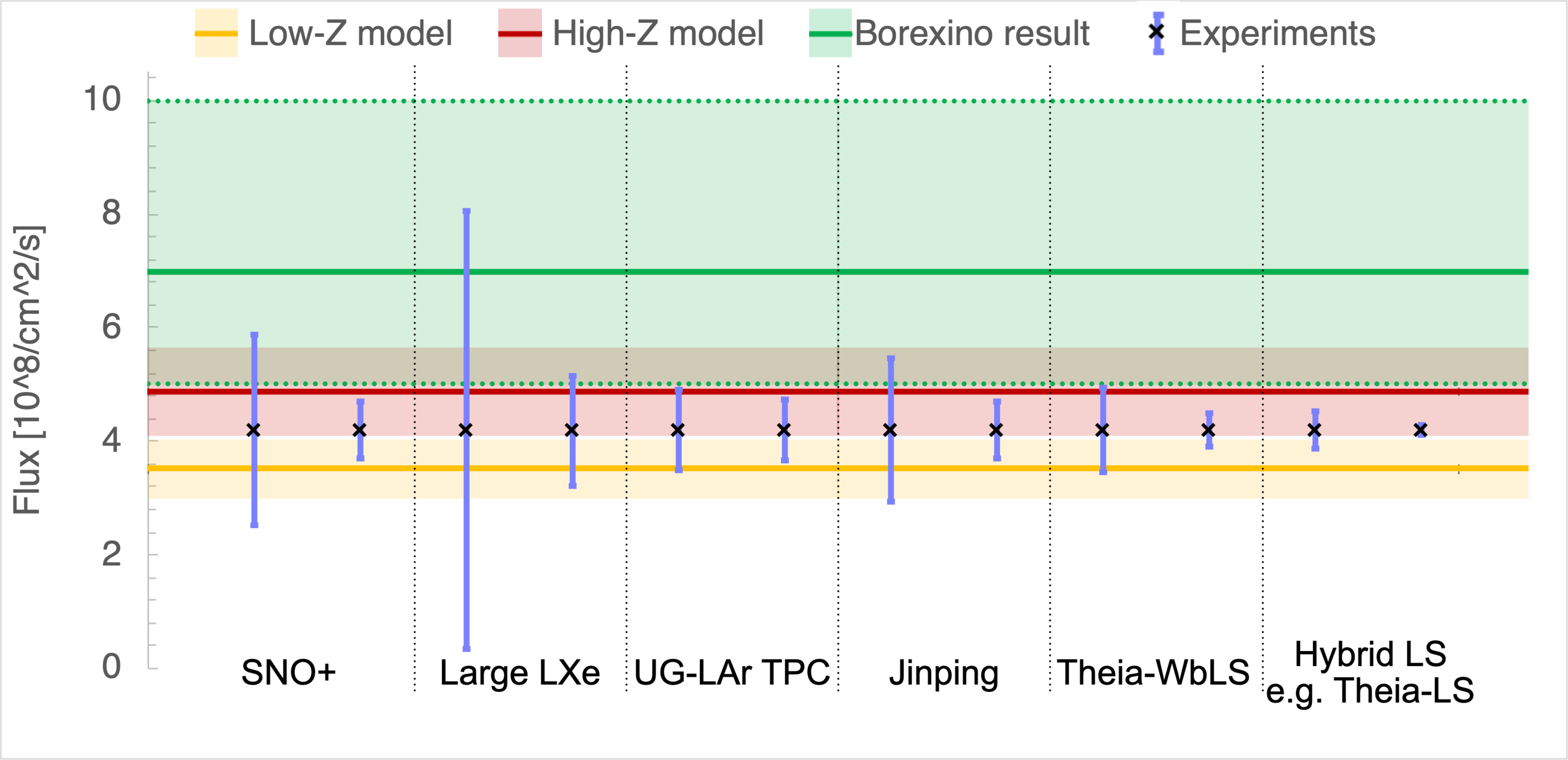}
    \caption{Prospects for measurement of the CNO neutrino flux, overlaid on the high- and low-Z model predictions, and the current observation from Borexino~\cite{2020Natur.587..577B}.  Shaded bands represent uncertainties.  Projected data points for each experiment are plotted at the midpoint between the high- and low-Z model predictions.  Two data points for each experiment encompass a range of possible detector scenarios, with the left point being the more conservative (scenario 1) and the right being the more aggressive (scenario 2) - as described in Table~\ref{t:cno}.  Data from~\cite{CNOgrandprix,LouisCNO,LArTPC,Andringa:2015tza,JinpingLOI,TheiaWP,MeVSolar}.}
    \label{f:cno}
\end{figure}

Figure~\ref{f:cno} shows an overview of potential future measurements of the CNO neutrino flux.  
Since the majority of these experiments are in the proposal stage, not yet funded, a degree of uncertainty exists in the projected detector capabilities and background levels.  Two possible scenarios are shown for each project, a nominal and a stretch goal, and Table~\ref{t:cno} describes the assumptions for each data point.  It is worth noting that the use of a 1.4\$ constraint on the pep flux is following the procedure adopted by Borexino in their CNO discovery paper~\cite{2020Natur.587..577B}, so allows for a like-for-like comparison.  Full details of backgrounds and detector assumptions can be found in the relevant references.

\begin{table}
\begin{tabular}{l|c|c}
Experiment & Scenario 1 & Scenario 2 \\
\hline\hline
SNO+~\cite{Andringa:2015tza} & \multicolumn{2}{c}{5 year exposure}\\
        & \multicolumn{2}{c}{1.4\% \emph{pep} constraint}\\
        & Current background levels & x10 reduction of $^{238}$U, $^{232}$Th \\
        &                           & x1000 reduction of $^{210}$Bi \\ \hline
Large LXe~\cite{LouisCNO} & 200 ton-yr exposure & 2000 ton-yr exposure \\
        & 10$^2$ reduction of 2$\nu\beta\beta$ & 10$^3$ reduction of 2$\nu\beta\beta$ \\
        & 1\% knowledge of backgrounds & Perfect knowledge of backgrounds \\ \hline
UG-LAr TPC~\cite{LArTPC} & \multicolumn{2}{c}{400 ton-yr exposure}\\
        &\multicolumn{2}{c}{6000 pe/MeV, 0.6MeV threshold}\\
        &\multicolumn{2}{c}{Underground argon (negligible impact from $^{42}$Ar, or pile up with $^{39}$Ar)}\\
        &$^{222}$Rn at 200$\mu$Bq/100 ton & $^{222}$Rn at 10$\mu$Bq/100 ton \\\hline
Jinping~\cite{JinpingLOI} & \multicolumn{2}{c}{1500 days exposure}\\
        & 1 ktonne fiducial   & 4 ktonne fiducial \\
        & 500 pe/MeV        & 1000 pe/MeV \\\hline
\textsc{Theia}-WbLS~\cite{TheiaWP}   & \multicolumn{2}{c}{5 years exposure} \\
        & 12.5ktonne fiducial & 60 ktonne fiducial \\
        & 55$^{\circ}$ angular resolution & 45$^{\circ}$ angular resolution\\\hline
Hybrid LS & \multicolumn{2}{c}{5 years exposure}\\
e.g. \textsc{Theia}-LS~\cite{MeVSolar}        & 500 tonne fiducial & 25 ktonne fiducial \\
        & ns-scale PMT TTS & 1.6ns or better TTS \\
        & \emph{or} 1.4\% constraint on pep & no constraint on pep \\

\end{tabular} 
\caption{Detector assumptions incorporated into Figure~\ref{f:cno}. \label{t:cno}}
\end{table}

With no single-purpose detectors on the horizon, it seems likely that the future solar neutrino program will rely heavily on multi-purpose detectors, from those designed for rare event searches, to long baseline neutrino experiments.  Such detectors offer a wealth of potential opportunities, on which we must capitalise in order to progress this field.  The percent-level precision desirable for the pp flux may be most likely to come from the large noble liquid detectors planned for dark matter searches, due to their low threshold, low intrinsic background, and good resolution.  
It would be interesting to be able to observe a larger number of pp-neutrinos per day in real time, which would allow the observer to follow these measurements for a significant period, to study their long term behaviour with reasonable statistics. 
MeV-scale measurements -- CNO and $^8$B spectral measurements -- may come from detectors designed primarily for NLDBD searches, such as large LS and hybrid detectors.